\documentclass[aps,prb,10pt,twocolumn,a4paper,superscriptaddress,nofootinbib, floatfix]{revtex4-2}

\usepackage{graphicx}
\usepackage{amsmath}
\usepackage{amssymb}
\usepackage{amsfonts}
\usepackage{color}
\usepackage{xcolor}
\usepackage[colorlinks = true,
            linkcolor = blue,
            urlcolor  = blue,
            citecolor = blue,
            anchorcolor = blue,
            unicode]{hyperref}

\DeclareMathOperator{\maxmin}{max(min)}

\begin{document}

\title{Diode effect in Shapiro steps in an asymmetric SQUID with a superconducting nanobridge} 

\author{Dmitrii~S.~Kalashnikov}
\email{kalashnikov.ds@phystech.edu}
\affiliation{Moscow Institute of Physics and Technology, 141700 Dolgoprudny, Russia}

\author{Gleb~S.~Seleznev}
\email{seleznev.gs@phystech.edu}
\affiliation{L.~D.\ Landau Institute for Theoretical Physics RAS, 142432 Chernogolovka, Russia}
\affiliation{Moscow Institute of Physics and Technology, 141700 Dolgoprudny, Russia}

\author{Andrei~Kudriashov}
\affiliation{Moscow Institute of Physics and Technology, 141700 Dolgoprudny, Russia}

\author{Ian~Babich}
\affiliation{Moscow Institute of Physics and Technology, 141700 Dolgoprudny, Russia}

\author{Denis~Yu.~Vodolazov}
\affiliation{Moscow Institute of Physics and Technology, 141700 Dolgoprudny, Russia}
\affiliation{Institute for Physics of Microstructures, Russian
Academy of Sciences, 603950, Nizhny Novgorod, GSP-105, Russia}

\author{Yakov~V.~Fominov}
\affiliation{L.~D.\ Landau Institute for Theoretical Physics RAS, 142432 Chernogolovka, Russia}
\affiliation{Moscow Institute of Physics and Technology, 141700 Dolgoprudny, Russia}
\affiliation{Laboratory for Condensed Matter Physics, HSE University, 101000 Moscow, Russia}

\author{Vasily~S.~Stolyarov}
\email{stolyarov.vs@phystech.edu}
\affiliation{Moscow Institute of Physics and Technology, 141700 Dolgoprudny, Russia}
\affiliation{National University of Science and Technology MISIS, 119049 Moscow, Russia}
\affiliation{Dukhov Research Institute of Automatics (VNIIA), 127055 Moscow, Russia}

\begin{abstract}

We investigate the Josephson diode effect in an asymmetric SQUID consisting of a sinusoidal Josephson junction formed by a Bi$_2$Te$_2$Se flake and a superconducting Nb nanobridge with a linear and multivalued current-phase relation (CPR). Current-voltage characteristics were measured both in the absence (\emph{dc} regime) and presence (\emph{ac} regime) of external microwave irradiation. Our \emph{dc} measurements reveal only weak critical current asymmetry (i.e. weak Josephson diode effect), while confirming the multivalued behavior of the SQUID. At the same time, the key finding of this work is the observation of strong Shapiro step asymmetry (concerning the \emph{dc} current direction) in the \emph{ac}   regime at finite magnetic flux. This peculiarity oscillates as a function of magnetic field with the SQUID’s periodicity and varies non-monotonically with the increase in microwave power. Our theoretical model shows that the pronounced Shapiro step asymmetry, despite the small diode effect in critical current, arises from the interplay between the sinusoidal and multivalued CPRs of the junctions.

\end{abstract}

\maketitle

\section{Introduction}\label{sec:Introduction}
Traditionally, a diode is an electronic device that allows current to flow in only one direction while blocking it in the opposite direction. Its semiconductor implementations, based on $p-n$ junctions, are widely used in various applications, making them essential components in modern electronics \cite{MalvinoBook}. Recently, there has been growing interest in superconducting diodes, driven by both fundamental research and potential applications in superconducting electronics \cite{Jiang2022, SDE_nadeem2023, Moll2023}.

The key characteristic of the superconducting diode effect (SDE) is the asymmetry of the critical current magnitude with respect to the current direction. There are various approaches to achieve this effect, which typically require broken time-reversal and inversion symmetries. This can be accomplished by introducing external magnetic fields in systems exhibiting spin-orbit interaction or by using topological and noncentrosymmetric materials 
\cite{Levitov1985, Edelstein1996, Wakatsuki2017, Yasuda2019, Ando2020, Daido2022.PhysRevLett.128.037001, Yuan2022, Ilic2022.PhysRevLett.128.177001, He2022, Kokkeler2022.PhysRevB.106.214504, Karabassov2022.PhysRevB.106.224509, Levichev2023.PhysRevB.108.094517, Kokkeler2024.10.21468/SciPostPhys.16.2.055, Mironov2024.PhysRevB.109.L220503, Hasan2024.PhysRevB.110.024508, Kochan2023arXiv, Mazanik2025arXiv}. {Moreover, the SDE can be caused by the vortex motion in systems with edge \cite{Vodolazov2005,Cerbu2013,Ustavschikov2022,Suri2022,Hou2023,Castellani2025} or bulk \cite{Villegas2003,Vondel2005,Plourde2008,Dobrovolskiy2020,Lyu2021} natural/artificial defects.}

One possible approach is to examine the SDE in systems that include Josephson junctions (JJs), where it is referred to as the Josephson diode effect (JDE) 
\cite{Yokoyama2014.PhysRevB.89.195407, Chen2018.PhysRevB.98.075430, Kopasov2021.PhysRevB.103.144520, Zhang2022.PhysRevX.12.041013, Davydova2022, Halterman2022.PhysRevB.105.104508, Baumgartner2022, Costa2023.PhysRevB.108.054522, Meyer2024.10.1063/5.0211491, Debnath2024.PhysRevB.109.174511, Chatterjee2024, Cuozzo2024, Sivakumar2024arXiv, Karabassov2024arXiv, Nikodem2024arXiv}.
The simplest setups for investigating the JDE are multiterminal systems and asymmetric SQUIDs using multiple sinusoidal Josephson junctions as their building blocks \cite{Haenel2022arXiv,Gupta2023, Bozkurt2023.10.21468/SciPostPhys.15.5.204}. 
These setups also provide a convenient way to tune both the strength and sign of the effect by varying the external magnetic fluxes through the loops. An alternative approach is to employ JJs with nonsinusoidal CPRs \cite{Fominov2022.PhysRevB.106.134514, Souto2022.PhysRevLett.129.267702, Ciaccia2023.PhysRevResearch.5.033131, Coraiola2024, Zhang2024.10.21468/SciPostPhys.16.1.030, Wu2025}. In this regard, superconducting nanobridges emerge as particularly promising candidates owing to their nearly linear, multivalued CPRs, which significantly distinguish them from ordinary sinusoidal junctions \cite{GolubovRevModPhys.76.411, LikharevRevModPhys.51.101, LikharevYakobson1975, Troeman2008, Giazzoto2021}. Additionally, these devices are characterized by their small size and easy fabrication process \cite{wernsdorfer2009micro, granata2016nano}. {However, despite extensive investigations of superconducting nanobridges \cite{Hasselbach2002, Hopkins2005Science, Dausy2021PhysRevApplied, Vondel2022}, particularly regarding the observation of the diode effect \cite{BezryadinAsymmetricSQUID, Margineda2023, Greco2023.10.1063/5.0165259}, the influence of the multivalued nature of the nanobridge CPR on the peculiarities of the JDE has not yet been explored.}

Another interesting approach to study the superconducting diode effect involves exploring quantities beyond just the asymmetry of critical currents. Although this phenomenon is the most prominent manifestation of the SDE, other parts of the current-voltage characteristics (CVC) also exhibit asymmetric behavior. For example, when capacitive effects are present, the JDE manifests as an asymmetry in the retrapping currents \cite{Misaki2021.PhysRevB.103.245302, Wu2022, Steiner2023.PhysRevLett.130.177002, SeleznevPhysRevB.110.104508}. Moreover, under external irradiation, features known as Shapiro steps appear in the CVC when the internal voltage oscillations of the system synchronize with the external driving alternating signal \cite{Likharev,Barone}. {The Shapiro steps can also demonstrate the JDE revealed in the asymmetry between shapes and sizes of the negative and positive steps (corresponding to different current directions) \cite{Fominov2022.PhysRevB.106.134514, Souto2022.PhysRevLett.129.267702, SeleznevPhysRevB.110.104508, Valentini2024, Li2024, Ciaccia2024, Leblanc2024.PhysRevResearch.6.033281}. However, the asymmetries mentioned above require specific conditions, such as the presence of effective capacitance of the junctions forming the SQUID (for the asymmetry of the retrapping currents) or the application of microwave irradiation (for the asymmetry of the Shapiro steps).} As a result, observing these effects is a more challenging task and has received less attention. At the same time, asymmetries in various parts of the CVC may reflect distinct features of the system (e.g., static and dynamic properties) and can exhibit entirely different behavior. Consequently, investigating the various asymmetric features of the CVC can provide additional valuable insights into the system.

In our work, we study an asymmetric SQUID composed of a sinusoidal Josephson junction and a superconducting nanobridge. Our focus is on the peculiarities of the JDE arising from the interplay between single-valued and multivalued Josephson junctions. We demonstrate that this interplay manifests itself in a strong asymmetry of the Shapiro features, which can be more pronounced than the asymmetry of the critical current.

The paper is organized as follows:
In Sec.~\ref{sec:Asymmetric SQUID}, we describe the experimental sample and the corresponding theoretical model for static superconducting properties of the asymmetric SQUID with a superconducting nanobridge.
In Sec.~\ref{sec:CVC}, we present experimental results and theoretical analysis of the SQUID current-voltage characteristics, measured both without (critical current measurements) and with (Shapiro steps measurements) microwave irradiation.
In Sec.~\ref{sec:Analysis of the Shapiro features}, we analyze the dependence of the Shapiro steps asymmetry on the magnetic flux and microwave irradiation power. 
In Sec.~\ref{sec:Discussions}, we discuss the observed asymmetric features and the physical mechanisms underlying them.
In Sec.~\ref{sec:Conclusions}, we present our conclusions.
Finally, additional details regarding calculations and auxiliary measurement results are provided in the Appendices.

\section{Asymmetric SQUID with superconducting nanobridge}
\label{sec:Asymmetric SQUID}
\subsection{Josephson junctions}\label{sec: JJs}

\begin{figure}[ht!]
\begin{center}
\includegraphics[width=8.5cm]{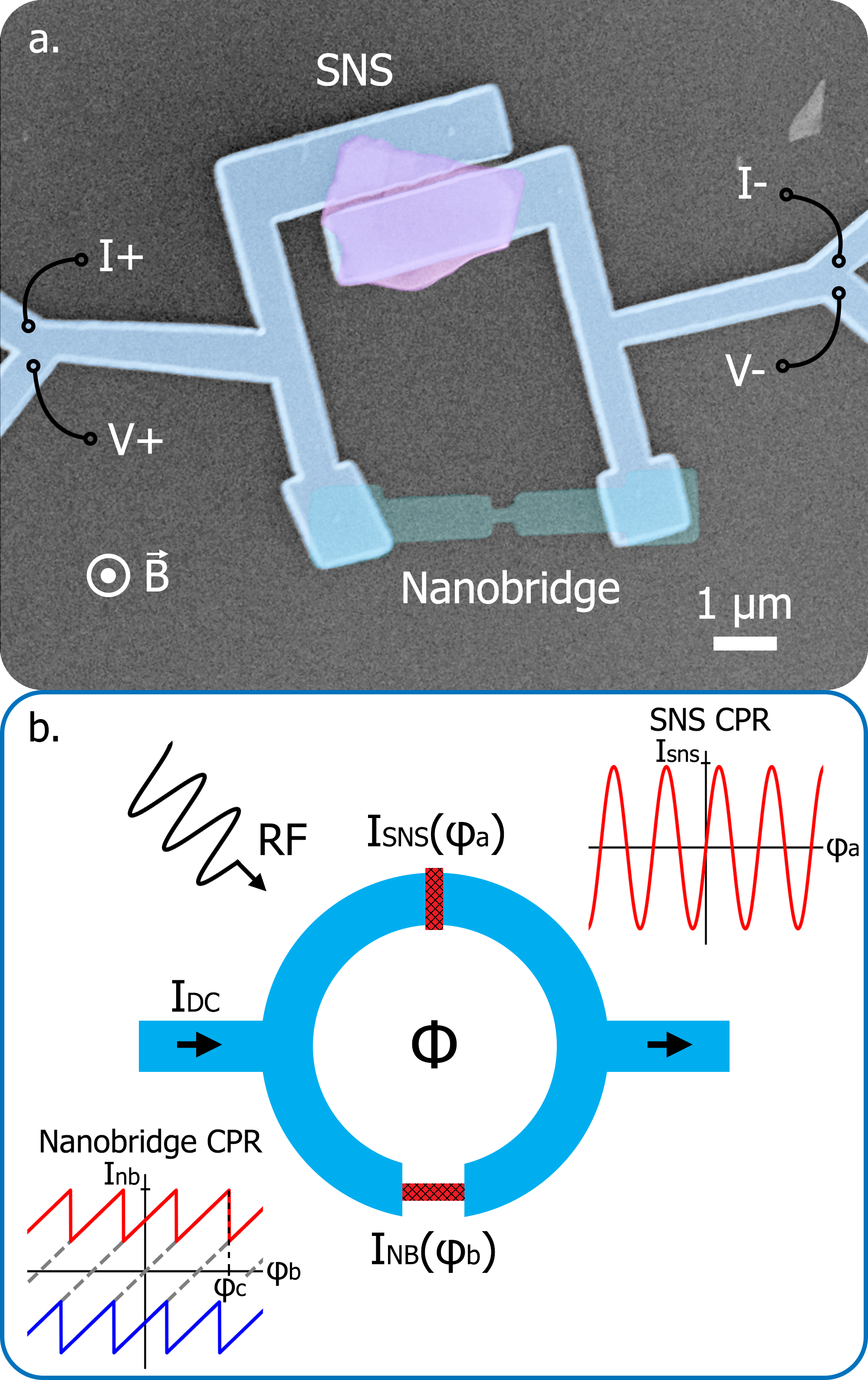}
\caption{Asymmetric SQUID structure. (a) False-colored Scanning Electron Microscopy (SEM) image of the sample. The top junction is a Superconductor/Normal metal/Superconductor (SNS) Josephson junction formed by a Bi$_2$Te$_2$Se flake, and the bottom one is a niobium nanobridge (NB). (b) An equivalent scheme of the SQUID with two different Josephson junctions. The insets show their current-phase relations (CPRs): sinusoidal function for the SNS junction and linear multivalued function for the nanobridge, where the critical current $I_{\mathrm{nb}}$ corresponds to the critical phase $\varphi_c > \pi$. In the nonstationary state with finite average voltage (positive or negative), the nanobridge CPR takes a sawtooth form, shown by red and blue segments, respectively. External radio-frequency ($rf$) irradiation can be applied to the SQUID.}
\label{fig: Image_and_scheme}
\end{center}
\end{figure}

In our work, we study the asymmetric SQUID, whose electrodes are made from thick niobium ($d \approx 100$\,nm) and whose arms contain two strongly different weak links. A SEM image of the SQUID is shown in Fig.~\ref{fig: Image_and_scheme}(a) and a simplified scheme of the system is presented in Fig.~\ref{fig: Image_and_scheme}(b).
The first junction of the SQUID is an SNS Josephson junction formed by a Bi${}_2$Te${}_2$Se flake with a thickness of about 90 nm. Similar samples using the same material were investigated in Ref.\ \cite{Kudriashov_Babich_2023}, where it was found that the flake acts as a normal metal and the junction has a standard sinusoidal current-phase relation $I_{\mathrm{SNS}}(\varphi_a) = I_{\mathrm{sns}} \sin{\varphi_a}$. The junction has large lateral dimensions, and its area $S_{\mathrm{sns}}$ is comparable to the area of the superconducting loop $S_{\mathrm{loop}}$. Therefore, when analyzing the SQUID, we account for the non zero magnetic flux through the junction in an applied magnetic field. As a result, the critical current of the SNS junction exhibits Fraunhofer modulation in the presence of a magnetic field, following $I_{\mathrm{sns}}(\Phi) = I_{\mathrm{sns}}\frac{\sin (\pi\Phi_{\mathrm{sns}}/\Phi_0)}{(\pi\Phi_{\mathrm{sns}}/\Phi_0)}$, where the flux $\Phi_{\mathrm{sns}}$ penetrating the SNS junction is proportional to the magnetic flux $\Phi$ through the superconducting loop as $\Phi_{\mathrm{sns}} = \Phi S_{\mathrm{sns}}/S_{\mathrm{loop}}$.

The second junction is a superconducting nanobridge made of a thin film of niobium with a thickness of 20\,nm, a width of 220\,nm, and a length of 380\,nm. Its CPR $I_{\mathrm{NB}}(\varphi_b)$ is assumed to be multivalued, which is typical for superconducting nanobridges with small cross sections \cite{GolubovRevModPhys.76.411, LikharevRevModPhys.51.101,Troeman2008,Giazzoto2021,Vondel2022,LikharevYakobson1975}. Generally, in the stationary regime (S state), the CPR of a JJ can be generated by a single branch defined in the interval [$-\varphi_{c}, \varphi_{c}$] and periodically translated by $2 \pi n$, where $n$ is an integer. A conventional single-valued CPR corresponds to $\varphi_{c} = \pi$. At the same time, in the nanobridge case, the generating branch is linear and $\varphi_{c}> \pi$. This leads to the multivaluedness of the CPR with $n$ enumerating its different branches. For the nanobridge, the critical phase also corresponds to the critical current $I_{\mathrm{nb}}$. In the nonstationary regime (R-state) with finite average voltage (positive or negative), we suppose that $I_{\mathrm{NB}}(\varphi_{b})$ effectively takes a sawtooth form \cite{Shmidt, Dinsmore2008fractional, Fujimaki, Trabaldo_2019}. Every time the phase reaches the value $\pm \varphi_c + 2 \pi n $, it switches to the neighboring branch, $n \mapsto n \pm 1$. The two signs correspond to the red and blue vertical segments in Fig.~\ref{fig: Image_and_scheme}(b), respectively.

This behavior of the nanobridge can also be understood in terms of vorticity defined as $N_v=\oint \nabla \varphi dl/2\pi$, where the integration is performed around the loop perimeter and $\varphi$ is the superconducting phase. Different branches of the CPR are distinguished by an integer $n$, which equals the vorticity $N_{v}$ at zero phase drop across the junction. As we increase the phase across the junction, fluxes can penetrate the SQUID loop, resulting in a change in the vorticity by one. Because of the nanobridge significant length, the phase gradient along it is small, leading to a correspondingly small superconducting velocity. Therefore, many vortices can enter the loop before the nanobridge reaches its critical current, which occurs at $N_{v} \gg 1$ or equivalently at $\varphi_{c} \gg \pi$ (for theoretical estimation of $\varphi_{c}$, see Appendix \ref{appendix: estimation of critical phase}). Moreover, once the phase reaches its critical value, a phase slip occurs in the nanobridge, causing a phase drop of $2\pi$. This is formally equivalent to changing $n$ by one, as mentioned earlier.

\subsection{Current-phase relation of the SQUID}\label{sec:CPR of the SQUID}
As two weak links are connected by a superconducting loop, the phase shift $\phi$ between them is related to the magnetic flux $\Phi$ through this loop as $\phi = \varphi_a - \varphi_b = 2\pi\Phi/\Phi_0$, where $\Phi_0$ is the magnetic flux quantum. We neglect the loop self-inductance, which is small in our case (see Appendix~\ref{appendix:geometric inductance}). 
At the same time, the supercurrent flowing through the SQUID is given by the sum of the junctions’ supercurrents,
\begin{equation}\label{eq:supercurrent}
I_{s}(\varphi) =I_{\mathrm{sns}} (\phi) \sin(\varphi + \phi) +  I_{\mathrm{NB}}(\varphi),
\end{equation} 
where we take into account the Fraunhofer modulation of the SNS critical current mentioned earlier. The asymmetry of the junctions forming the SQUID leads to the appearance of the JDE in our system.

\begin{figure*}[ht!]
\begin{center}
\includegraphics[width=17.5cm]{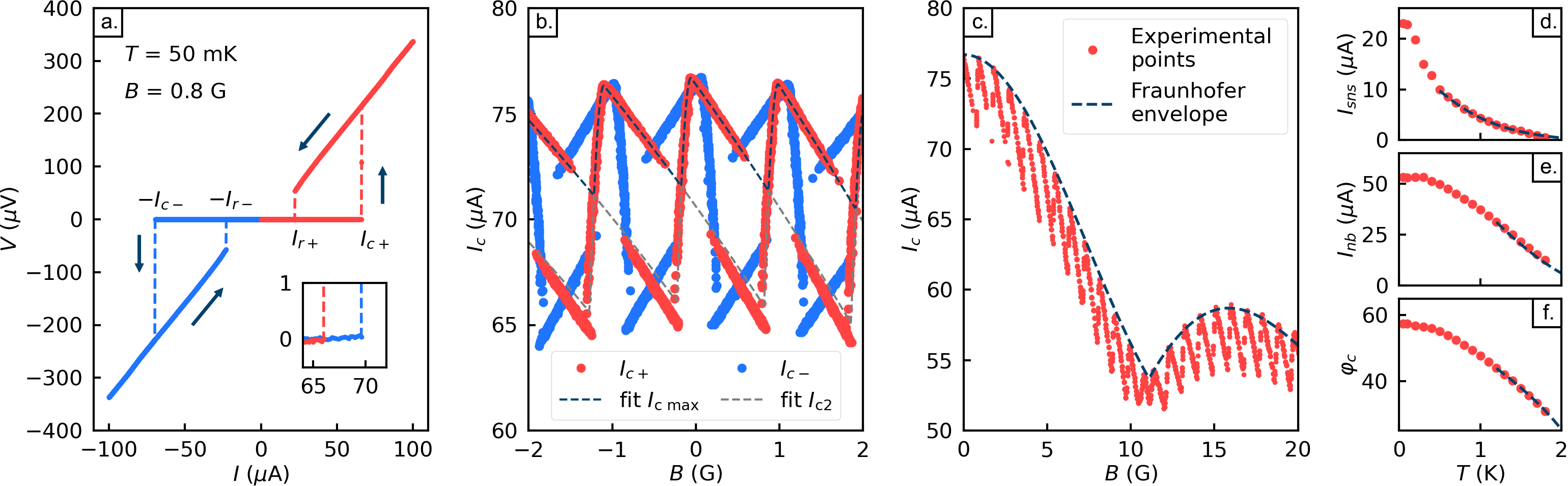}
\caption{Direct current measurements of the sample. (a) CVC of the SQUID measured at temperature $T = 50$\,mK and in an external magnetic field $B = 0.8$\,G. The black arrows indicate the current sweep direction. The hysteretic behavior of the curve is associated with local overheating. The inset shows a zoomed area of the IV curve around the critical currents in absolute values of current and voltage. The critical currents differ in opposite directions by around 4\,$\mu$A. (b) Dependence of $I_{c}$ of the SQUID on the external magnetic field in positive and negative current directions. The critical currents in both directions slightly differ from each other, which is a manifestation of the diode effect. A set of CVCs was measured at a fixed value of magnetic field, and the dots indicate the critical current obtained from each CVC. As can be seen, at a given magnetic field, the dots concentrate around several values of the critical current, demonstrating the multivalued CPR of the SQUID. Different critical currents correspond to switching to the resistive state from different branches of the CPR. 
{The dashed lines represent fits to the two branches of the multivalued $I_{c}(B)$ curve for the positive critical currents with Eqs.\ \eqref{eq: sharp drops} and \eqref{eq: smooth segments}. The line labeled $I_{\text{c max}}$ corresponds to the} {highest branch (with the largest critical current value), while $I_{\text{c2}}$ represents the lower (second) possible critical current branch at a fixed magnetic field.} (c) Dependence of the highest branch of the positive critical current over a wide range of fields. The small-scale oscillations of $I_c (B)$ (because of the flux through the whole SQUID) are modulated by the Fraunhofer dependence of the critical current through the SNS junction, shown by the dashed line. The right panels demonstrate the temperature dependence of (d) the SNS critical current $I_{\mathrm{sns}}$, (e) nanobridge critical current $I_{\mathrm{nb}}$ and (f) critical phase $\varphi_c$. Theoretical fits at high temperature are shown by dashed lines. 
}
\label{fig: Ic(H)}
\end{center}
\end{figure*}

To highlight the unique features associated with the presence of a superconducting nanobridge, we express the supercurrent of the SQUID in the R state as a Fourier series
\begin{equation} \label{eq:CPR in non stat regime}
\frac{I_{s}(\varphi)}{I_{\mathrm{nb}}} = A_{0\pm} + \sum_{k=1}^{\infty} A_{k\pm} \sin \bigl(k (\varphi +  \delta_{k\pm}) \bigl),
\end{equation}
where $\pm$ sign indicates positive and negative part of the sawtooth CPR (red and blue, respectively).
The zeroth harmonics $A_{0\pm} = \pm (\varphi_{c} - \pi)/\varphi_{c}$ arise from the nanobridge CPR and determine the average supercurrent. The amplitudes $A_{k\pm}$ and phases $\delta_{k\pm}$ at $k=1$ are given by the following equations:
\begin{gather}\label{eq: A1pm}
 A_{1 \pm} = \sqrt{\left( \frac{ I_{\mathrm{sns}}(\phi) }{ I_{\mathrm{nb}} }\right)^2 + \frac{4}{\varphi_c^2} -  \frac{4 I_{\mathrm{sns}} (\phi) }{ I_{\mathrm{nb}}  \varphi_c } \cos{ (\phi \pm \varphi_{c}} )} , \\
 \tan \left(\delta_{1 \pm} \pm \varphi_{c} \right) =   \frac{\sin{ (\phi \pm \varphi_{c}) } }{ \cos{ (\phi \pm \varphi_{c}) } - 2 I_{\mathrm{nb}} /\varphi_{c} I_{\mathrm{sns}} (\phi) } \label{eq: phases},
\end{gather}
while $A_{k \pm} = -2 /k \varphi_c$ and  $\delta_{k\pm} = \mp \varphi_{c}$ for $k > 1$.

As shown by Eqs.\ \eqref{eq:CPR in non stat regime}--\eqref{eq: phases}, there are two different mechanisms leading to the diode effect. The first is the amplitude asymmetry of the first harmonic ${A_{1+} \neq A_{1-}}$ at $\sin \varphi_{c} \sin \phi \neq 0$. Visually, $\sin \varphi_{c} \neq 0$ corresponds to the horizontal offset between red and blue sawtooth patterns in Fig.~\ref{fig: Image_and_scheme}(b).  The second mechanism involves phase shifts between the first and higher Josephson harmonics, where $\delta_{1\pm} \neq \delta_{k\pm}$ at least for some $k > 1$. While the phase-shift mechanism of the JDE in SQUIDs is well known \cite{Fominov2022.PhysRevB.106.134514, Souto2022.PhysRevLett.129.267702}, the amplitude-asymmetry mechanism has not been previously reported to our knowledge. This asymmetry stems from the interference between the sinusoidal CPR and the first harmonic of the nanobridge sawtooth CPR. Notably, this effect is not present in SQUIDs that consist solely of junctions with single-valued CPRs because of trivial values of the critical phase, $\sin \varphi_c = 0$. Therefore, it highlights the multivalued nature of the nanobridge CPR. As we demonstrate below, the $A_{1+} \neq A_{1-}$ asymmetry drives the pronounced Shapiro steps asymmetry observed in the nonstationary regime.

\section{Current-voltage characteristics}
\label{sec:CVC}
We begin the discussion of the experimental results with the analysis of the CVC of the SQUID. Figure~\ref{fig: Ic(H)}(a) shows a typical example of an IV curve observed in the experiment. As long as the voltage is zero, the SQUID is in the S state. When the applied current exceeds a critical value $I_c$, the voltage changes abruptly, the CVC becomes linear, and the SQUID switches into the R state. When the current decreases back, the voltage remains finite even for current values less than the critical value. It drops back to zero only at a smaller value $I_r$, manifesting hysteretic behavior of the system. This hysteretic behavior is associated with local overheating of the sample caused by Joule dissipation  \cite{hazra2010hysteresis, blois2016heat} and not by capacitive effects. {In particular, the temperature effectively changes from $T = 50$~mK in the S state (fixed base refrigerator temperature) to $T \sim 2$~K near $I_{r}$ in the R state. This temperature is estimated from the condition that the transition from the R to S state corresponds to the equality between the $T$-dependent critical current and the applied dc current.}

\subsection{Critical current measurements} \label{sec: critical current measurements}

\begin{figure*}[ht!]
\begin{center}
\includegraphics[width=17.5cm]{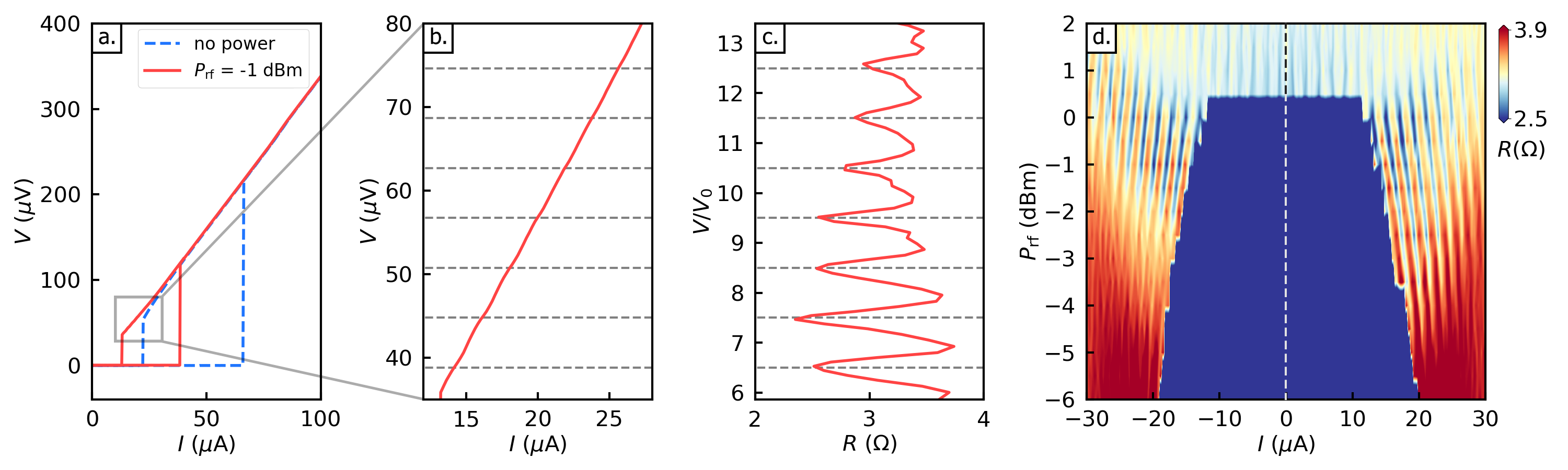}
\caption{SQUID measurements under radio-frequency (rf) irradiation at $\Phi = 3\Phi_0/4$.
(a) Comparison between IV curves of the SQUID with applied microwave irradiation at frequency $f_{\text{rf}} = 2.888$\,GHz (solid red) and without it (blue dashed line). (b) Zoomed area of the backward branch of the IV curve near $I_r$ with applied microwave irradiation. The lines bend slightly at the voltages $V_n = (n-1/2)V_0$ with integer $n$ and $V_0 = \Phi_0 f_{\text{rf}}$, shown by dashed gray lines. (c) The same data as in panel (b) but with the horizontal axis representing the calculated differential resistance $R = dV/dI$. The voltage is now expressed in units of $V_0$. (d) Dependence of the differential resistance on the backward branches vs current through the sample and rf power near $I_{r\pm}$.
}
\label{fig: IV_RF}
\end{center}
\end{figure*}

To investigate the diode effect in the critical currents, we conducted \emph{dc} measurements at finite magnetic fields. The IV curve in Fig.~\ref{fig: Ic(H)}(a) was measured at an external magnetic field of 0.8\,G. The inset shows that the critical currents in the positive ($I_{c+}$) and negative ($I_{c-}$) directions differ by approximately 4\,$\mu$A.

For a detailed analysis of this asymmetry, we measured the critical current dependence on magnetic field over a wide range. The low-field results are shown in Fig.~\ref{fig: Ic(H)}(b). We measured 100 CVCs at given magnetic field values to obtain critical current statistics. The procedure of collecting statistics is important in our case because the transition to the R state may occur at different currents at the same magnetic field, see Fig.~\ref{fig: Ic(H)}(b). This effect arises from the existence of different vorticity states at the same value of magnetic field, which is a manifestation of the multivalued nature of the nanobridge CPR. Similar effects were previously observed in asymmetric SQUIDs with two different nanobridges \cite{Vondel2022, BezryadinAsymmetricSQUID}.  At the same time, the results for higher magnetic fields are presented in Fig.~\ref{fig: Ic(H)}(c). In this case, magnetic fluxes penetrate the SNS junction, inducing Fraunhofer-like modulation of SQUID oscillations.

{We fit the $I_{c}(B)$ dependence in Figs.~\ref{fig: Ic(H)}(b) and~(c) simultaneously maximizing the absolute value of the supercurrent given by Eq.\ \eqref{eq:supercurrent}. As demonstrated earlier in Ref.\ \cite{Kudriashov_Babich_2023}, the phase values at which critical currents are reached can be either $ \pm \varphi_{c}$ or $\varphi^{*}_{\pm} =  \pm 2 \pi N_{v\pm} \pm \arccos{\left(-\frac{I_{\mathrm{nb}}  }{\varphi_c I_{\mathrm{sns}}(\phi)}\right)} - \phi$, which correspond to the maximum/minimum positions (numerated by integer vorticity numbers $N_{v\pm}$) within the range $[-\varphi_c, \varphi_c]$ (for more details, see Appendix \ref{appendix:asymmetry of the critical currents}).}

{When critical currents are reached at the $\pm \varphi_{c}$, they take the following values
\begin{equation}\label{eq: sharp drops}
    I_{c\pm} = I_{\mathrm{nb}} + I_{\mathrm{sns}}(\phi) \sin (\varphi_{c} \pm \phi),
\end{equation}
which corresponds to the steep (almost vertical) segments in the $I_{c}(B)$ dependence in Fig.~\ref{fig: Ic(H)}(b).}

{Conversely, when critical currents are reached at $\varphi^{*}_{\pm}$, they are given by the expression 
\begin{multline}\label{eq: smooth segments}
    I_{c\pm} = I_{\mathrm{sns}}(\phi) \sqrt{1 -\left(\frac{I_{\mathrm{nb}}}{\varphi_{c} I_{\mathrm{sns}}(\phi)}\right)^2} \\ +\frac{I_{\mathrm{nb}}}{\varphi_{c}} \left[ 2 \pi N_{v\pm} + \arccos{\left(-\frac{I_{\mathrm{nb}}  }{\varphi_c I_{\mathrm{sns}}(\phi)}\right)} \mp \phi\right],
\end{multline}
which corresponds to the not-so-steep linear segments in Fig.~\ref{fig: Ic(H)}(b). 
According to this expression, there are multiple critical current values at the fixed magnetic field, distinguished by the vorticity number $N_{v\pm}$. In Fig.~\ref{fig: Ic(H)}(b) $I_{\text{c max}}$ corresponds to the} {highest branch (with the largest critical current value), while $I_{\text{c}2}$ represents the lower (second) critical current branch. The $I_{\text{c}2}$ branch does not align perfectly with the experimental data, likely because of a slight deviation of the nanobridge CPR from ideal linearity assumed in the theory.}

{Nevertheless, our theoretical model successfully reproduces the overall shapes of the experimentally measured $I_{c}(B)$ curves. At the same time, we emphasize that our theory does not account for the apparent ``discontinuities'' in the $I_{c}(B)$ dependence, which manifest themselves in the absence of experimental points in some segments of the experimental curve. These discontinuities are attributed to stochastic switching from different vorticity states to the R state and the metastability of these vorticity states, as previously discussed in  Ref. \cite{Dausy2021PhysRevApplied}. However, these effects are not the focus of our research}.

To evaluate the strength of the diode effect {in the dc regime}, we calculate the diode efficiency, $\eta = |(I_{c+}-I_{c-})|/(I_{c+}+I_{c-})$, considering only branches with the largest $I_{c\pm}$ values at fixed magnetic field. In our system, the diode efficiency reaches its maximum value of $\eta = 2.6\%$, which corresponds to $\approx 4$\,$\mu$A difference. Hence, our measurements reveal only weak asymmetry in the critical currents despite the high asymmetry of the junctions. This can be explained by vorticity. Indeed, critical currents of the SQUID are reached in the vicinities of $\pm \varphi_{c}$, which corresponds to {$N_{v\pm} \gg 1$} vortices trapped inside the loop. At the same time, vorticities differ by one at $I_{c \pm}(B)$. Therefore, the difference in critical currents is approximately $\Delta I_{c} \sim I_\mathrm{nb} (2\pi/\varphi_c)$ which is close to the experimental value of 4\,$\mu$A and small because of the large value of $\varphi_{c}$ [see Fig.~\ref{fig: Ic(H)}(f)].

{Finally, in addition to studying the critical current asymmetry, we applied the fitting procedure described above to the set of measurements at different temperatures. This allows us to extract the temperature dependencies} of the main parameters of the system, which are shown in Figs.~\ref{fig: Ic(H)}(d)--\ref{fig: Ic(H)}(f) (for more details, see Appendix \ref{appendix:temperature dependence of junction parameters}). The critical currents of both JJs vanish at temperatures significantly lower than the critical temperature of the niobium electrodes, which is approximately 8\,K. We relate the low critical temperature of the nanobridge to its small thickness (20\,nm) and the fabrication technique: The film was produced using the thermal sputtering method, resulting in numerous defects and high oxidation of the niobium. At the same time, for the SNS junction, we attribute the reduced critical temperature to the fact that the Josephson coupling occurs not directly between the niobium electrodes. Instead, it emerges between flake regions beneath the electrodes, where only weak proximity-induced superconductivity is present. This effect was previously investigated in Ref.\ \cite{stolyarov2020josephson} for similar flakes.

\begin{figure*}[ht!]
\begin{center}
\includegraphics[width=17.5cm]{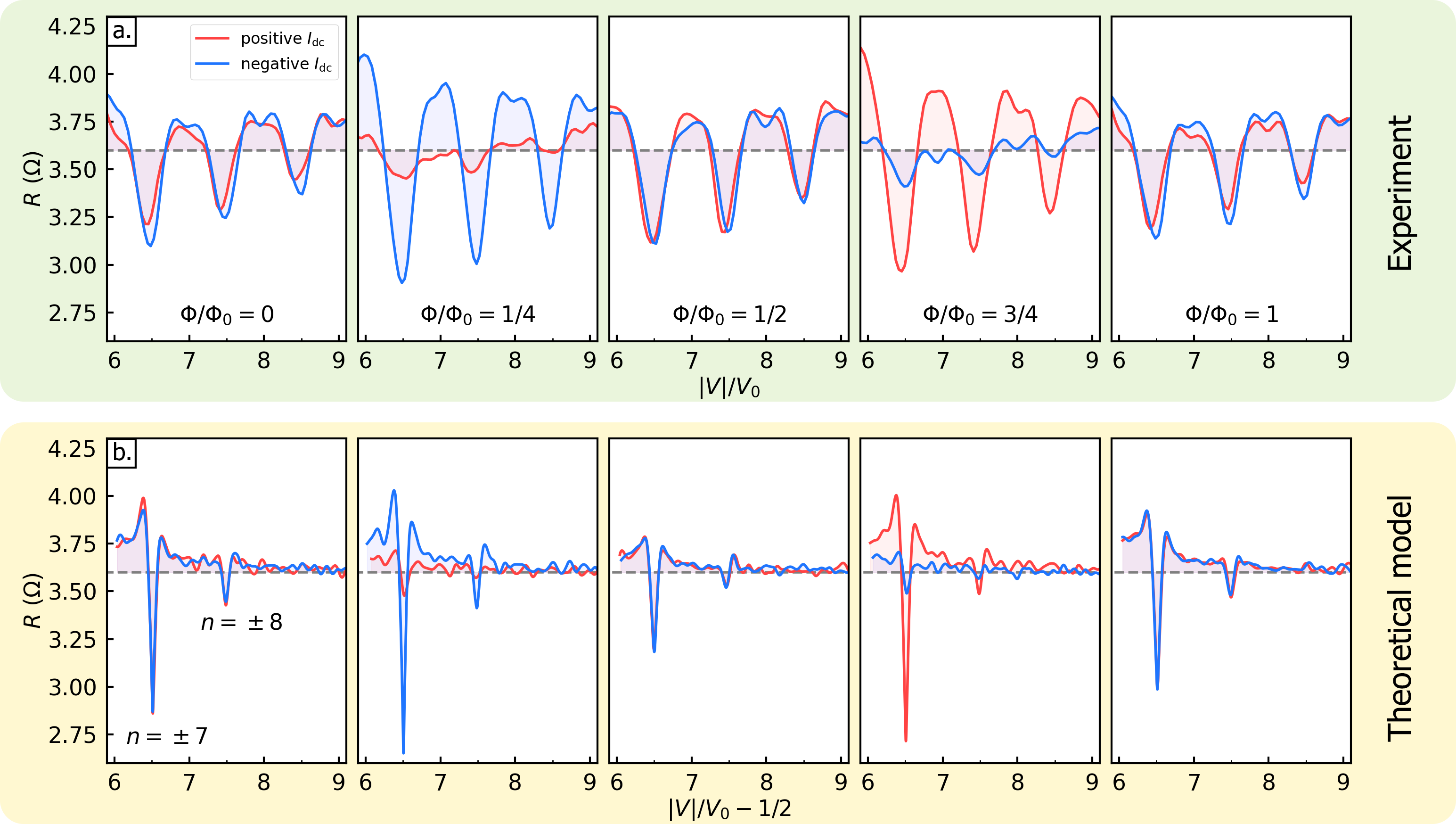}
\caption{Demonstration of the diode effect in Shapiro features (thermally smeared steps) in magnetic field, which reveal themselves as minima of $R = dV/dI$. (a) Five plots of the differential resistance $R$ dependence as a function of voltage at different magnetic fields. The data are obtained by numerically differentiating experimentally measured IV curves with applied microwave irradiation at $f_{\mathrm{rf}} = 3.754$\,GHz and power $P_{\mathrm{rf}} = 1$\,mW. The red and blue lines correspond to the positive and negative current $I_{\text{dc}}$ through the SQUID, respectively. The dashed gray lines indicate the reference value $R_0$. (b) Results of the theoretical calculations within the RSJ model [Eqs.\ \eqref{eq:J1beg} and \eqref{eq:J2beg}] with $T = 1.8$\,K, $\varphi_{c} = 30$, $I_{\mathrm{sns}} = 0.3\,\mu$A, $I_{\mathrm{nb}} = 12\,\mu$A, $I_{\mathrm{ac}} = 12\,\mu$A. The theory qualitatively reproduces the experimentally observed features. However, the horizontal axis in the theoretical subplots had to be manually shifted by $1/2$, see Appendix \ref{appendix:Shapiro steps shift} for discussion. 
}
\label{fig: Shapiro_5}
\end{center}
\end{figure*}

\subsection{Shapiro steps measurements} \label{sec:Shapiro measurments}

In addition to measuring the CVC without external irradiation, we also explore the unique effects that occur when microwave irradiation is applied to the sample. {We conducted our experiments at different frequencies in the 2.8--3.8\,GHz range and observed qualitatively similar behavior.} Figure~\ref{fig: IV_RF} shows an example of a change in the SQUID IV curve under such conditions with rf frequency $f_{\text{rf}}=2.888$\,GHz. As expected, the critical current decreases, since an alternating current is added to the direct current. {However, Shapiro steps are not formed on the forward current sweep branch; instead, the system jumps directly into a normal state. One reason of is the large self-heating effect, which suppresses superconducting properties. The second reason is the large voltage values [see red and blue vertical segments that lie outside the gray box in Fig. \ref{fig: IV_RF}(a)], which correspond to the large numbers of Shapiro features characterized by small sizes and strong thermal smearing, making them unobservable in the experiment. Therefore, we take a closer look at the reverse R branch near $I_r$ shown in Fig.~\ref{fig: IV_RF}(b) where the effective temperature and voltage values are significantly lower}. In this part of the CVC, we observe synchronization of the JJs with external electromagnetic irradiation similar to those explored in Refs.\ \cite{Dinsmore2008fractional, Nanobridges_withRF_Shelly2020}. However, in our measurements, the observed peculiarities on the IV curve do not form clear steps but rather smeared features, which are caused by thermal fluctuations. That is why we address them as Shapiro features instead of Shapiro steps. This effect is clearly visible in the differential resistance dependence, plotted in  Fig.~\ref{fig: IV_RF}(c), which demonstrates periodic dips of about 30\% of the maximum value. These features occur at periodic intervals of $V_0 = 6.0\,\mu$V, corresponding to the frequency of the applied irradiation $f_\mathrm{rf} = 2.888$\,GHz. However, all stripes are located at voltages $V_0 (n - 1/2)$ with integer $n$ demonstrating a shift of $V_0/2$ from the expected positions (for the discussion of possible reasons, see Appendix \ref{appendix:Shapiro steps shift}).

The evolution of the differential resistance with rf power is shown in  Fig.~\ref{fig: IV_RF}(d). The dark blue region corresponds to the S state of the SQUID. Its border represents the critical current $I_{r}$ of the backward branch of the CVC, which monotonically decreases with RF power. The blue stripes in the R state correspond to the Shapiro features that change their positions along the current axis, but remain at the same voltages throughout the entire power range. Another notable feature is the abrupt disappearance of the critical currents $I_{r\pm}$ as a function of $P_{\mathrm{rf}}$ [upper border of blue region in Fig~\ref{fig: IV_RF}(d)]. In this region,  where $P_{\mathrm{rf}} \gtrsim 0.5$~dBm, the superconducting properties are still present, which are manifested by the Shapiro features [see Fig.~\ref{fig: Appendix_moredata}(f) below]. Similar peculiarities were previously observed in Refs.\ \cite{Cecco2016,Biswas2018, Nanobridges_withRF_Shelly2020} where they were attributed to the heating effect caused by microwave irradiation. However, in our research, we do not focus on this region of $P_{\mathrm{rf}}$.

To investigate the JDE in the Shapiro features, we measure them in different magnetic fields. Figure~\ref{fig: Shapiro_5}(a) shows the evolution of differential resistance as the magnetic flux through the SQUID changes. At zero field, the $R_{\text{diff}}(V)$ dependencies are nearly identical for both current directions. Their minor difference is caused by a slight deviation of the magnetic field from zero in the experiment. With increasing field, the $R(V)$ dips on the positive current branch become smaller in amplitude, while on the negative branch they increase. Their asymmetry reaches a maximum at $\Phi = \Phi_0/4$. Then, the process is reversed, and at $\Phi_0/2$ the curves coincide again. At $3\Phi_0/4$, the difference reaches the maximum again but with the opposite diode effect sign. Finally, at $\Phi_0$ the state of the SQUID becomes indistinguishable from the state at zero flux, so the curves coincide again. 

In this way, we demonstrate that the amplitude of differential resistance dips differs between the positive and negative current directions, which is one of the manifestations of the JDE. The diode effect reaches its maximum at magnetic fluxes $\Phi \approx \Phi_{0}/4$ and $\Phi \approx 3\Phi_{0}/4$, where the Shapiro features nearly vanish for one current direction while remaining distinctly visible for the opposite direction. This observation highlights a strong asymmetry in the Shapiro features, contrasting with the relatively weak asymmetry observed in the critical currents measured in the absence of microwave irradiation.

To confirm that the observed asymmetry of the Shapiro steps is caused by the asymmetric structure of the SQUID, we numerically simulate the phase dynamics using the RSJ model \cite{Barone, Likharev} with the supercurrent given by Eq.\ \eqref{eq:supercurrent},
\begin{gather} \label{eq:J1beg}
d \varphi/d \tau + I_{s}(\varphi)/I_{\mathrm{nb}} = j_{\mathrm{dc}} + j_{\mathrm{ac}} \cos{\omega \tau} + \xi(\tau),\\
\label{eq:J2beg}
v = d \varphi/d \tau,
\end{gather}
where we define the dimensionless variables: $\tau = \omega_{J} t$, $v = V/I_\mathrm{nb} R$, $j_{\mathrm{dc}} = I_{\mathrm{dc}}/I_\mathrm{nb}$, $j_{\mathrm{ac}} = I_\mathrm{ac}/I_\mathrm{nb}$, $\xi = I_{f}/I_\mathrm{nb}$, and  $\omega = \Omega/ \omega_{J}$. Here, $I_{\mathrm{dc}}$ and $I_{\mathrm{ac}}$ represent the \emph{dc} and \emph{ac} currents, respectively, $I_{f}$ is the fluctuating thermal current, $\Omega$ is the frequency of the \emph{ac} current, $R_0 = 3.6$ \,$\Omega$ is the normal resistance taken from the experiment (see Table~\ref{table: samples} for sample N7 which is the main object of our study), and  $\omega_{J} = (2e/\hbar) I_{\mathrm{nb}} R_0$ is the Josephson frequency. The thermal fluctuations in Eq.\ \eqref{eq:J1beg} are considered as white noise with correlator $\langle \xi(\tau)\xi(\tau') \rangle = 2 {\cal{T}} \delta(\tau - \tau')$ and dimensionless temperature ${\cal{T}} = 2eT/\hbar I_\mathrm{nb}$. While the temperature generally varies along the IV curve because of the Joule heating, it can be treated as constant within narrow intervals of $V/V_0$. 

As a result, we obtain the CVC, $\overline{v}(j_{\mathrm{dc}})$ dependence, where $\overline{\dots\vphantom{v}}$ indicates time-averaging. The comparison of theoretical results with experimental data is presented in Fig.~\ref{fig: Shapiro_5}. {While full numerical agreement has not been achieved, our model qualitatively describes the asymmetric dependence of the Shapiro features on magnetic flux. However,  a notable inconsistency can be observed: the experimental Shapiro steps appear to be less influenced by thermal fluctuations, exhibiting a slower decrease in amplitude with increasing voltage than our theory predicts. One potential explanation for this discrepancy might be the overestimation of the effective temperature used in our numerical simulations. Nonetheless, additional results obtained at lower temperatures, presented in Appendix \ref{appendix: comparison of the results for high and low temperatures}, exhibit the same issue and do not align quantitatively with the experimental data.} {Therefore, this mismatch between theory and experiment might be attributed to the non-equilibrium processes in the electronic subsystem that cannot be characterized by a single effective temperature parameter $T$ and lie beyond the theoretical framework of the RSJ model.}

\section{Analysis of the Shapiro features} \label{sec:Analysis of the Shapiro features}

\begin{figure*}[ht!]
\begin{center}
\includegraphics[width=17.5cm]{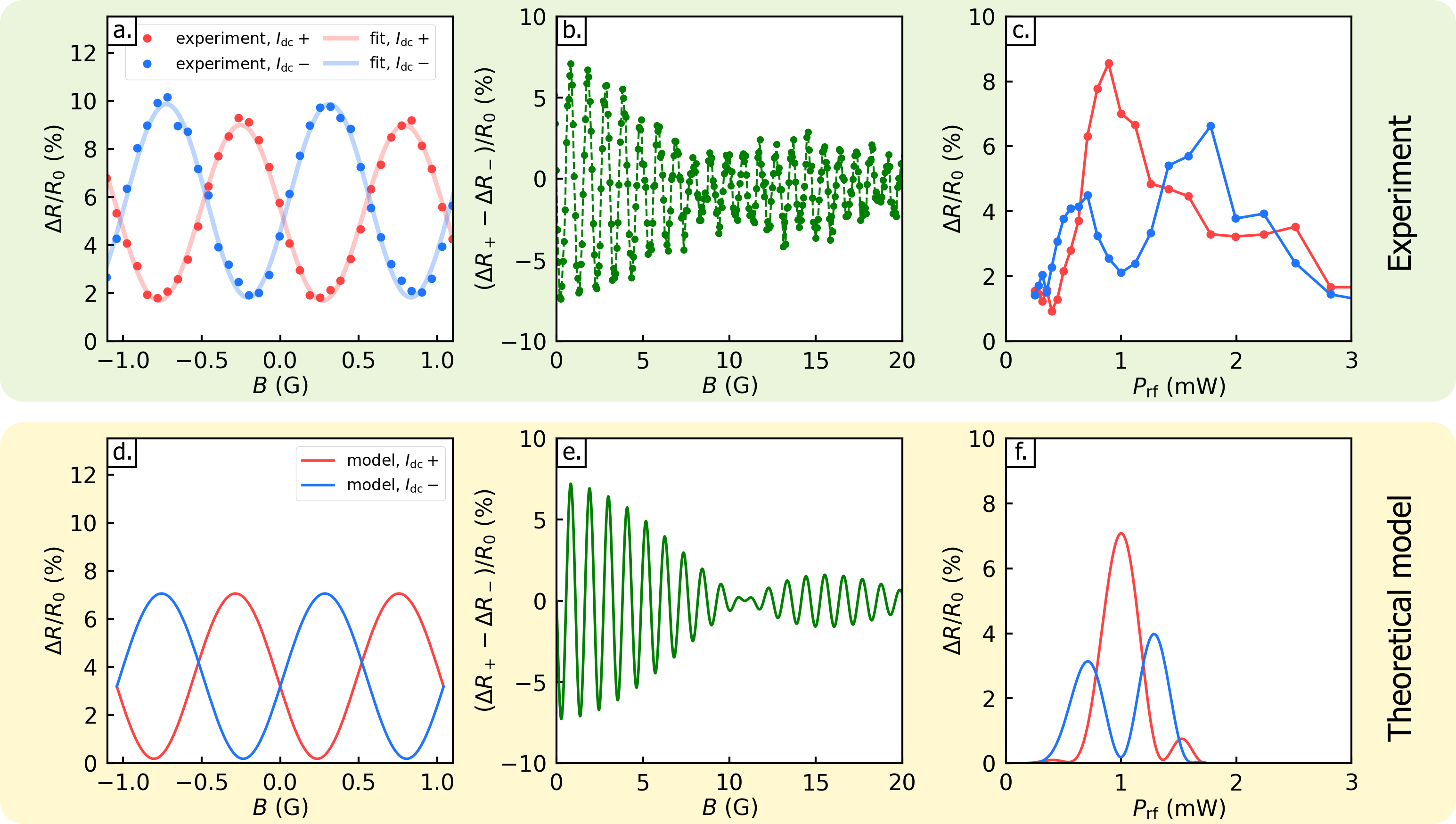}
\caption{Quantitative analysis of the Shapiro features and comparison between the experimental data, shown in the top row, and the theoretical results [obtained from Eq.\ \eqref{eq:R_diff}], shown in the bottom. (a) and (d) Dependence of the Shapiro features depths $\Delta R_{\pm}$ on the magnetic field at $|V|/V_0 = 6.5$ (corresponding to $n = \pm 7$ in the theoretical model). The dots represent experimental data for positive (red) and negative (blue) current directions, collected with an applied microwave signal at a frequency $f_{\text{rf}} = 3.754$\,GHz and a generator power $P_{\text{rf}} = 1$\,mW. The solid lines indicate fits with a sinusoidal function, which show good agreement with the experimental points. The theoretical results are also close to a sinusoidal function. (b) and (e) Fraunhofer envelope of the Shapiro features depths asymmetry $\Delta R_{+} - \Delta R_{-}$ over a wide range of the fields. The oscillations of the diode effect strength with SQUID periodicity at small magnetic fields and Fraunhofer envelope at large magnetic fields demonstrate that the asymmetry of the Shapiro features arises from the interplay of the two junctions. (c) and (f) Dependence of $\Delta R_{\pm}$ on microwave power at a fixed magnetic field $B = 0.8$\,G, which corresponds to a flux $\Phi/\Phi_0 \approx 0.77$. The curves exhibit a nonmonotonic dependence with periodic changes in the sign of the diode effect. We attribute this behavior to the heating effect of microwave irradiation, which manifests itself through the temperature dependence of $\varphi_{c}$. The crossings of the curves correspond to the condition $\sin \varphi_{c}(T) = 0$.}
\label{fig: Shapiro_analysis}
\end{center}
\end{figure*}

For a more detailed analysis of the Shapiro feature asymmetry, we examine its variation with magnetic field and rf power. We quantify the $n$th Shapiro feature depths (for both positive and negative integers) using the relative change in differential resistance at the feature centers $R_{\pm}$ from the value in the normal state: $\Delta R_{\pm} = (R_{0} - R_{\pm})$ (for more details about extracting these values from the experimental data, see Appendix~\ref{sec: Experimental methods}). For definiteness, we focus on the Shapiro feature with $n = \pm 7$ (corresponds to $|V|/V_0 = 6.5$ in the experiment). At the same time, features with other numbers demonstrate similar behavior.

The obtained experimental result for the Shapiro features dependence on the magnetic field is shown in Fig.~\ref{fig: Shapiro_analysis}(a). The curves for both current directions exhibit a periodic dependence on the field and can be well fitted by a sinusoidal function, which is plotted in the same figure with solid lines. The oscillation periods obtained during the fit for the positive and negative branches are $\Delta B_+ = \Delta B_- = 1.04 $~G, which exactly coincide with the SQUID period measured by the critical current oscillations, see Fig.~\ref{fig: Ic(H)}(b).
Furthermore, we investigate the dependence of the Shapiro features asymmetry, $ \Delta R_{+} - \Delta R_{-} $, over a wider range of magnetic field, as demonstrated in Fig.~\ref{fig: Shapiro_analysis}(b). At larger magnetic field values, the envelope of the oscillations in the asymmetry of the Shapiro features has the same form as the Fraunhofer dependence of the SQUID critical currents $I_{c\pm}$ on the magnetic field, see Fig.~\ref{fig: Ic(H)}(c). These observations demonstrate that the diode effect in the Shapiro features arises because of the asymmetric structure of the SQUID.

At the same time, the study of the asymmetry of the Shapiro features reveals their non-monotonic dependence on the power of the microwave signal. As demonstrated in Fig.~\ref{fig: Shapiro_analysis}(c), at a fixed magnetic flux value through the SQUID, $\Phi = 3\Phi_0/4$, with an increase in the power of the rf signal, the depths of the Shapiro features alternately reach their maxima in the positive and negative current directions. However, this behavior is radically different from the Bessel-law dependence typically observed for the amplitudes of the Shapiro features in similar experiments with JJs.

To explain these results, we develop a theoretical framework using the ``Slowly Varying Phase'' perturbation method \cite{StephenPhysRev.186.393, Kautz1996, Likharev} concerning the smallness of the supercurrent and thermal fluctuations compared to the \emph{dc} current. In addition, we focus on the vicinity of the centers of the Shapiro features. In this case, we separate the phase dynamics into ``fast'' [varying on a short time scale $\tau \lesssim (n \omega)^{-1}$] and ``slow'' [varying on a long time scale $\tau  \gg (n \omega)^{-1}$] parts, and then conduct averaging over an intermediate timescale $(n \omega)^{-1}\ll \Delta \tau \ll {\cal{T}}^{-1}, (\hat{v} - n \omega)^{-1}$. We use the ``hat'' symbol for functions averaged over such timescales. This procedure allows us to exclude fast dynamics and derive the effective (Josephson) equations (for more details see Appendix\ \ref{appendix: effective Josephson equations}) for slow functions,
\begin{gather}\label{eq:effective}
    \dot{\theta}_{n}(\tau) + \hat{I}_{s\pm}(\tau) =  \Delta j + \xi(\tau),
\end{gather}
where $\dot{\theta}_{n}$ and $\Delta j$ are deviations of the average (over time) voltage  and current from the centers of the $n$th Shapiro features: 
\begin{gather}\label{eq:slow variables}
    \hat{v}(\tau) = n \omega  + \dot{\theta}_{n}(\tau),  \quad j_{\mathrm{dc}} =   n \omega + A_{0\pm} + \Delta j,
\end{gather}
and $\hat{I}_{s\pm}(\tau)$ are the averaged supercurrents in the R state (excluding the $A_{0\pm}$ term, which determines the location of the Shapiro feature) for positive and negative CVC branches, respectively. The time dependence of $\hat{I}_{s\pm}(\tau)$ enters through the effective current-phase relations $\hat{I}_{s\pm}(\theta_{n})$. Equation \eqref{eq:effective} can be solved numerically.

The effective current-phase relations $\hat{I}_{s\pm}(\theta_{n})$ differ from the sawtooth CPR discussed previously in Sec.\ \ref{sec:CPR of the SQUID}. Their Fourier harmonics contain additional factors given by Bessel functions, see Eq.\ \eqref{eq:SC}. As a result of this, the amplitudes decrease faster than in the case of the sawtooth CPR. A reasonable approximation is then the ``single-harmonic approximation'', where only the first harmonic in the effective CPR is taken into account,
\begin{gather} \label{eq:Is pm} 
\hat{I}_{s\pm}(\tau) =  A_{1\pm} J_{n}\left(j_{\mathrm{ac}}/\omega \right)\sin\bigl( \theta_{n}(\tau) \bigl).
\end{gather}
 As we demonstrate below, this approximation provides a qualitatively correct description of the diode effect in our system because it captures the main mechanism, $A_{1-} \neq A_{1+}$ asymmetry.

The effective equation on the phase $\theta_{n}$ [Eq.\ \eqref{eq:effective}] with supercurrent given by Eq.\ \eqref{eq:Is pm} takes the form of the standard Josephson equation for a sinusoidal junction (however, with different critical currents for opposite current directions). Such equations have been studied in detail in many theoretical works \cite{ Ambegaokar1969.PhysRevLett.22.1364,FalcoSeries,Bishop, Ivanchenko1968}, allowing us to directly apply the well-known expressions for the CVC to our case (for more details, see Appendix \ref{sec: CVC in the single harmonic approximation}). As a result, we obtain the differential resistance at the centers of the Shapiro features:
\begin{equation}\label{eq:R_diff}
\frac{R_{\pm}}{R_0} = I^{-2}_0 \left(1/{{\cal{T}}_{\pm}}\right) = \begin{cases}
  \left(2 \pi  /{{\cal{T}}_{\pm}} \right) \exp({-2/{\cal{T}}_{\pm}}), \quad {{\cal{T}}_{\pm}} \ll 1, \\
 (1 - 1/2 {{\cal{T}}^2_{\pm}}), \quad {{\cal{T}}_{\pm}} \gg 1,
\end{cases}
\end{equation}
where $I_{0}$ is the modified Bessel function and  ${\cal{T}}_{\pm} = {\cal{T}}/A_{1\pm} J_n(j_{\mathrm{ac}}/\omega)$ is the effective dimensionless temperature that determines the strength of thermal fluctuation in different current directions near the centers of the $n$th Shapiro features.

The dependence of the Shapiro features on the magnetic field obtained from the theoretical model is shown in Figs.~\ref{fig: Shapiro_analysis}(d) and (e). 
The curves in Fig.~\ref{fig: Shapiro_analysis}(d), originating from the exact Bessel function $I_0$ in Eq.\ \eqref{eq:R_diff}, demonstrate an almost harmonic dependence on the magnetic field. 
Analytically, the dependence of this type arises in the high-temperature limit of Eq.\ \eqref{eq:R_diff}, $\Delta R_{\pm}/R_{0} =A_{1\pm}^2 J^2_{n}(j_{\mathrm{ac}}/\omega)/2 {\cal{T}}^2 \propto \mathrm{const} + I_{\mathrm{sns}}(\phi)\cos{(\phi \pm \varphi_c)}$, see Eq.\ \eqref{eq: A1pm}. At the same time, the $I_{\mathrm{sns}}(\phi)$ dependence manifests itself at larger magnetic fields, leading to the Fraunhofer envelope of the Shapiro features asymmetry, see Fig.~\ref{fig: Shapiro_analysis}(e). 
 
Theoretical results for the dependence of the Shapiro features' depths on the power of \emph{ac} irradiation are shown in Fig.~\ref{fig: Shapiro_analysis}(f). In our consideration, we take into account both effects of external irradiation, which are the generation of the \emph{ac} current $j_{\mathrm{ac}}$ and heating of the sample (for more details, see Appendix \ref{appendix: dependence on Pac}). Our findings indicate that it is the heating effect responsible for the observed changes in the sign of the diode effect. This occurs because the critical phase $\varphi_c$ depends on temperature, as shown in Fig.~\ref{fig: Ic(H)}(f). As the power of irradiation increases, it raises the temperature, which in turn changes the sign of $\sin \varphi_c$ with the corresponding change in the diode effect sign.

\section{Discussion}
\label{sec:Discussions}
\subsection{{Manifestation of the multivalued CPR in the Shapiro steps asymmetry}}
In the above consideration, we have primarily focused on the asymmetry of the Shapiro features, which occur under applied external microwave irradiation. As shown in Fig.~\ref{fig: Shapiro_5}, this asymmetry can be pronounced at specific values of the magnetic fluxes, $\Phi \approx \Phi_0/4$ and $\Phi \approx 3\Phi_0/4$. Theoretically, we explain this effect as arising from the asymmetry in the amplitudes of the first harmonics $A_{1+} \neq A_{1-}$ at $\sin \phi \sin \varphi_c \neq 0$. Confirmation of the dependence of this asymmetry on $\sin \varphi_c$ is evident in the relationship between the strength of the diode effect and the power of microwave irradiation [see Fig.~\ref{fig: Shapiro_analysis}(c)]. The point is that the temperature increase caused by irradiation effectively changes $\varphi_{c}$ [see Fig.~\ref{fig: Ic(H)}(f)], leading to periodic changes in the sign of the diode effect. This is a manifestation of the multivalued nature of the diode effect in the R state. In contrast, in SQUIDs solely consisting of JJs with single-valued CPRs, this mechanism of the diode effect does not exist. 

However, even for single-valued junctions, it is possible to observe asymmetry in the Shapiro steps \cite{Valentini2024, Li2024, Ciaccia2024, Leblanc2024.PhysRevResearch.6.033281}. In this case, the diode effect is caused by phase shifts between different harmonics in the effective SQUID current-phase relation \cite{Fominov2022.PhysRevB.106.134514, Souto2022.PhysRevLett.129.267702}. It is important to note that this mechanism of the Shapiro steps asymmetry differs from the one described above. We also emphasise that in previous works, asymmetries in only a few of the first Shapiro steps were investigated, while we observe up to ten asymmetric Shapiro features.

 \subsection{{Comparison of asymmetries in the critical current and Shapiro features}}
 
Additionally, we would like to address the coexistence of the weak diode effect in critical currents alongside the noticeable asymmetry observed in the Shapiro features. In the S state, as mentioned in Sec.~\ref{sec: critical current measurements}, the asymmetry of the critical currents reaches a maximum value $\Delta I_{c} = 2\pi I_{\mathrm{nb}}/\varphi_{c}$. At the same time, the diode efficiency $\eta \approx \pi I_{\mathrm{nb}}/\varphi_{c}(I_{\mathrm{nb}}+I_{\mathrm{sns}}) \approx 3\% $ is small because of the small critical current asymmetry compared to the large critical current values that are mostly determined by the zeroth harmonic contributions $A_{0\pm}$ [see Eq.\ \eqref{eq:CPR in non stat regime}]. This is a consequence of the large $\varphi_{c} = 58$ value at $T = 50$\,mK. 

However, in the R state where Shapiro features are measured, the temperature is high, $T \sim (1-3)$\,K, in contrast to the S state. In this regime, the critical currents are strongly suppressed, see Figs.~\ref{fig: Ic(H)}(d) and \ref{fig: Ic(H)}(e). This explains the strong smearing, which leads to small depths of the Shapiro features observed in the experiment. At the same time, the Shapiro steps' asymmetry is determined by the $A_{1\pm}$ harmonics [see Eq.\ \eqref{eq: A1pm}], which depend on the relation between $I_{\mathrm{sns}}$  and the amplitude of the first harmonic of the sawtooth nanobridge CPR,  $I^{(\mathrm{nb})}_{1} = 2I_{\mathrm{nb}} /\varphi_{c}$. In the R state, they are comparable to each other. For example, $I_{\mathrm{sns}} \approx 0.3$\,$\mu$A and $I^{(\mathrm{nb})}_{1} \approx 0.8$\,$\mu $A at $T = 1.8$\,K. This leads to the fact that the relative asymmetry between $A_{1+}$ and $A_{1-}$ is pronounced, resulting in strong asymmetry of the Shapiro features despite their small sizes. To summarize, the asymmetry of the critical currents is suppressed by the large $\varphi_{c}$ value, while the asymmetry of the Shapiro features is strong because of the pronounced relative asymmetry between $A_{1+}$ and $A_{1-}$.

\subsection{{Optimization of the critical phase value to enhance the diode effect}}

 {As follows from the above discussion, to enhance the difference in the SQUID critical current for positive and negative directions and use it as a Josephson diode operating in the dc regime, it is essential to decrease the critical phase value. At the same time, to reach a high relative asymmetry between $A_{1+}$  and  $A_{1-}$ in order to observe significant asymmetry in the Shapiro features in the ac regime, it is necessary to adjust the critical phase value so that $I_{\mathrm{sns}}$ and  $I^{(\mathrm{nb})}_{1}$  become comparable. This may involve both increasing and decreasing the critical phase value. Note that this parameter  is directly proportional to the length of the nanobridge, thus it can be controlled by selecting an appropriate nanobridge size. Alternatively, the critical phase value can be reduced by increasing the effective temperature, achievable through adjustments to the cryostat temperature or by applying external irradiation.} 

\subsection{{Increasing the visibility of the Shapiro steps}}
{
In our experiment, we observe smeared Shapiro steps, manifested as dips in the differential resistance of up to 30\%. This behavior is attributed to thermal fluctuations arising from the self-heating effects. The impact of these thermal fluctuations on the system dynamics can be quantitatively described using the effective normalized temperature $\cal{T}$ [see Eq. \eqref{eq:J1beg}].}
{Several strategies can be proposed to reduce this parameter and thereby enhance the visibility of the Shapiro features. One approach is to increase the critical currents of both junctions, which would suppress the relative influence of thermal fluctuations. For instance, Ref.  \cite{Nanobridges_withRF_Shelly2020} demonstrates clear Shapiro steps (a complete drop in resistance to zero) for a single nanobridge with critical current an order of magnitude larger than in our system. Another strategy involves mitigating the self-heating effects, which are responsible for the high effective temperature on the backward branch of the SQUID's CVC. One potential solution is to improve heat dissipation by using a substrate with higher thermal conductivity at cryogenic temperatures than the Si/SiO${}_2$ used in our setup --- for example, non-oxidized silicon or sapphire. Additionally, thermal hysteresis in superconducting nanobridges can be reduced by depositing a normal metal layer to enhance heat transfer \cite{Blois2013}. Finally, the introduction of a shunt resistance across the device can also be employed as an effective strategy to reduce the dissipation.}
\subsection{{Selecting a junction accompanying the nanobridge in the SQUID}}

{In this work, the Shapiro diode effect was investigated in the SQUID containing a Josephson junction formed through a Bi${}_2$Te${}_2$Se flake. This sample was selected based on our previous research \cite{Kudriashov_Babich_2023}, which demonstrates that, despite the specificity of the material, such a junction exhibits a standard sinusoidal CPR --- a key requirement for our model. At the same time, a sinusoidal CPR is also typical for other types of Josephson junctions that use metals or insulators as barriers \cite{Barone}. Consequently, our results are broadly applicable to such systems, with no fundamental constraints on the sinusoidal junction type. We emphasize that achieving high diode efficiency requires the sinusoidal junction critical current to be comparable to the first harmonic of the nanobridge sawtooth CPR.}

\subsection{{Limitations of the theoretical model}}
Finally, we note that some factors and features corresponding to the complex behavior of the system are not captured by our RSJ model, particularly the Shapiro features shift. In addition, there may be other mechanisms of temperature relaxation besides the electron-phonon mechanism assumed in our consideration. Taking them into account might increase the widths of the curves in Fig.~\ref{fig: Shapiro_analysis}(f), which are currently smaller than the experimentally observed ones. {Furthermore, the non-equilibrium processes, which cannot be described by the effective temperature $T$, might also occur in the system, potentially explaining the discrepancy between experiment and theory in Fig.~\ref{fig: Shapiro_5}, where the experimental Shapiro steps exhibit a more gradual decrease with voltage than the theory predicts.} At the same time, our relatively simple model qualitatively reproduces all the peculiarities observed in the experiment and relates them to the properties of the junctions. Hence, we expect that complicating the model will improve the results quantitatively but will not lead to qualitatively new features of the diode effect in the system.

\section{Conclusions}
\label{sec:Conclusions}

We have experimentally investigated the Josephson diode effect in the asymmetric SQUID containing a sinusoidal SNS junction and a superconducting nanobridge with a multivalued, almost linear current-phase relation. The multivalued nature of the current-phase relation of the nanobridge manifests itself in the nontrivial value of the critical phase  $\varphi_{c} \gg \pi$
which distinguishes it from conventional single-valued junctions. 

We conducted both \emph{dc} and \emph{ac} measurements of the SQUID current-voltage characteristics. While the \emph{dc} measurements revealed only weak critical current asymmetry, they confirmed the multivalued nature of the SQUID behavior and allowed us to extract the temperature dependence of the junction parameters. At the same time, in the \emph{ac} measurements, we observed strong asymmetry of the Shapiro features (for different current directions) with feature numbers up to 10 on the backward part of the Joule-heating-driven hysteretic CVC.

We investigated the dependence of the Shapiro features on the magnetic field for both current directions. In this case, we observed periodic oscillations of the Shapiro diode effect strength at low magnetic fields and a Fraunhofer envelope at higher magnetic fields. We also conducted measurements of the Shapiro features asymmetry concerning the power of external microwave irradiation $P_{\mathrm{rf}}$. These measurements demonstrated nonmonotonic behavior of the Shapiro features, which differs significantly from the expected Bessel-law dependence. All the above-mentioned dependencies show switching of the diode effect sign.

To explain the observed peculiarities, we developed a theoretical framework using the method of slowly varying phase and performed numerical simulations within the RSJ model with thermal noise. The results have demonstrated qualitatively good agreement between the theoretical predictions and experimental data.

The key observation of our theoretical model is that in our asymmetric SQUID at finite magnetic fields, the amplitudes of the Josephson harmonics generally differ for opposite current directions ($A_{1+} \neq A_{1-}$ in our case), leading to strong asymmetry of the Shapiro features. This effect requires a nontrivial value of the critical phase $\sin \varphi_c \neq 0$ and, hence, is not present in SQUIDs where both junctions are single-valued.

The sign and strength of the diode effect in the Shapiro features are controlled by the $\sin{\phi} \sin {\varphi_c} $ combination. Therefore, the periodic dependence of Shapiro features asymmetry on the magnetic field corresponds to the SQUID oscillations, while their nonmonotonic behavior concerning the power of microwave irradiation results from the heating effect and the corresponding temperature dependence of the critical phase $\varphi_c (T)$.

 \acknowledgments
We thank M. Yu. Kupriyanov for useful discussions. This work was supported by the Ministry of Science and Higher Education of the Russian Federation (Grant No. 075-15-2025-010, sample fabrication; Grant No. FSMG-2023-0014, data analysis and approximations) and by the Russian Science Foundation (Grant No. 25-42-00058, low-temperature measurements). G.S.S. acknowledges support from the Ministry of Science and Higher Education of the Russian Federation (Project No. FFWR-2024-0017). Ya.V.F. was supported by the Basic Research Program of HSE.

\appendix

\section{Estimation of \texorpdfstring{$\varphi_{c}$}{varphi.c}}
\label{appendix: estimation of critical phase}
We estimate $\varphi_c$ assuming that the critical current of the niobium nanobridge $I_{\mathrm{nb}}$ corresponds to the depairing current. The critical phase can be expressed as $\varphi_{c} = \varphi_{\mathrm{np}} + \varphi_{b}$, where $\varphi_{\mathrm{np}}$ is the phase drop along the narrow part of the nanobridge and $\varphi_{b}$ is the phase drop along its banks. Our nanobridge is long (in units of coherence length in niobium, $\xi_{\mathrm{Nb}} \sim 10$~nm) and diffusive. In this case, the Usadel model is applicable and predicts that the phase gradient in the nanobridge is equal to  $0.68/\xi_{\mathrm{Nb}}$ (see Fig.~2 in Ref.\ \cite{Romijn1982}) resulting in  $\varphi_{\mathrm{np}}=0.68l_{\mathrm{np}}/\xi_{\mathrm{Nb}} \approx 26$,  where $l_{\mathrm{np}} = 0.38$\,$\mu$m is the length of the narrow part. For $\varphi_{b}$ calculation, we use the fact that the phase drop along a segment at a fixed current behaves as $\varphi_{\mathrm{seg}} \propto l_{\mathrm{seg}}/w_{\mathrm{seg}}$, where $l_{\mathrm{seg}}$ and $w_{\mathrm{seg}}$ are the segment length and width, respectively. This relates the phase drop along the banks to the phase drop along the narrow part as $\varphi_{b} = \varphi_{\mathrm{np}}  (l_{b} w_{\mathrm{np}}/l_{\mathrm{np}} w_{b})$,  where $l_{b}=3.12$\,$\mu$m is the overall banks length, $w_{b}=0.7$\,$\mu$m and $w_{\mathrm{np}}=0.22$\,$\mu$m are the widths of the banks and the narrow part of the nanobridge, respectively. As a result, we obtain an estimate of the critical phase $\varphi_c= \varphi_{\mathrm{np}}(1 + l_{b} w_{\mathrm{np}}/l_{\mathrm{np}} w_{b} )  \approx 93$. 

Generally,  $I_{\mathrm{nb}}$ can be smaller than the depairing current, which would lead to a smaller value of the critical phase compared to the above estimate. This could explain the difference between the value obtained above and that determined from the direct fitting of the experimental data, see Fig.~\ref{fig: Ic(H)}(f). 

\section{Geometric inductances}
\label{appendix:geometric inductance}

To ensure that the applied magnetic flux $\Phi$ almost precisely coincides with the magnetic flux through the SQUID loop, we estimate the geometric inductances of the SQUID arms.

Generally, the total magnetic flux through the loop is the sum of the external flux $\Phi$ and the flux generated by the circulating current $I_{\text{circ}}$, which is given by $ L_{\text{loop}}I_{\text{circ}}$, where $L_{\text{loop}}$ is the geometric inductance. Using electromagnetic modeling with the sample geometry, we calculated $L_{\text{loop}} = 10$\,pH. The inductance of each SQUID arm can be estimated as $L_{\text{arm}} = L_{\text{loop}}/2 = 5$\,pH, which turns out to be much smaller than the inductances of the SNS junction  $L_{\text{sns}} = \Phi_0/2\pi I_{\text{sns}} = 14$\,pH and the nanobridge $L_{\text{nb}} = \varphi_c\Phi_0/2\pi I_{\text{nb}} = 340$\,pH calculated at the lowest temperature $T = 50$\,mK. In addition,  when we discuss Shapiro steps, we take into account that they appear on the overheated backward branch of the CVC, which leads to a significant decrease in critical currents. This means that the inductances of the JJs become several times larger. Therefore, in our model, we neglect the effect of the geometric inductance and assume that the flux in the SQUID equals the external flux $\Phi$.

\section{Asymmetry of the critical currents}\label{appendix:asymmetry of the critical currents}

The critical currents of the SQUID in different current directions are given by the extrema of supercurrent defined by Eq.\  \eqref{eq:supercurrent}:
\begin{equation}
    I_{c\pm} =  \left| \maxmin_{\varphi}\left(I_{\mathrm{sns}}(\phi) \sin (\varphi + \phi) + \frac{I_{\mathrm{nb}} \varphi}{\varphi_c}\right)\right|,
\label{eq: theory Ic(H)}
\end{equation}
where $\varphi$ lies within the range $[-\varphi_{c}, \varphi_{c}]$. {As discussed in Sec.\ref{sec: critical current measurements}, these extrema may occur either at the boundaries $ \pm \varphi_{c}$ or at the positions $\varphi^{*}_{\pm} =  \pm 2 \pi N_{v\pm} \pm \arccos{\left(-\frac{I_{\mathrm{nb}}  }{\varphi_c I_{\mathrm{sns}}(\phi)}\right)} - \phi$ numerated by integer vorticity numbers $N_{v\pm}$.}

\begin{figure*}[ht!]
\begin{center}
\includegraphics[width=17.5cm]{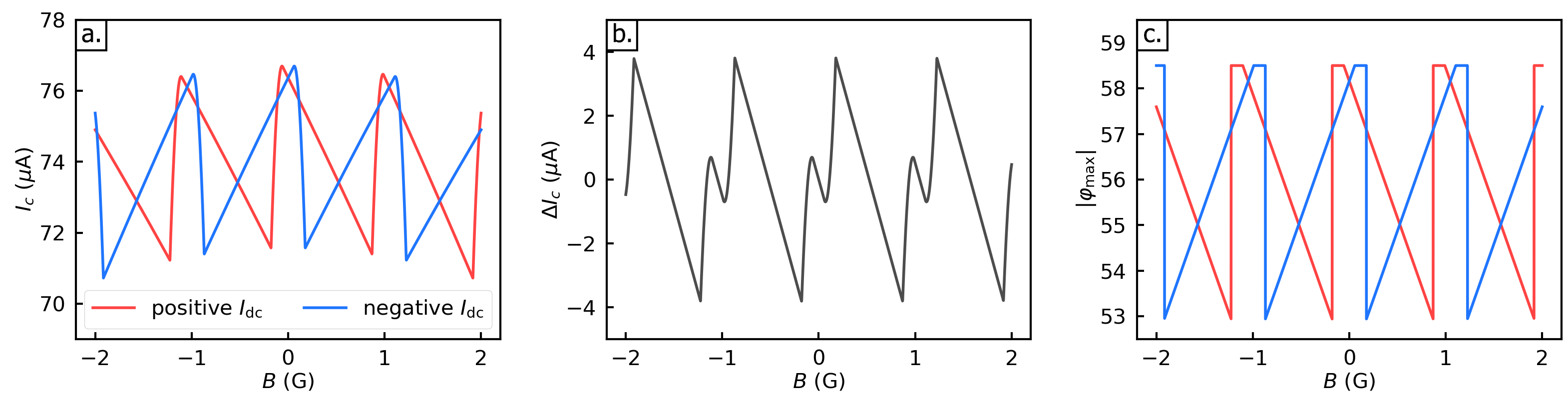}
\caption{Asymmetry of the critical currents in the \emph{dc} regime (theory). (a)~Reproduced theoretical curves from Fig.~\ref{fig: Ic(H)}(b) for the {highest branches} of the critical current dependencies on the magnetic field $B$ through the loop for positive (red) and negative (blue) current directions. In this scale, parts of sinusoidal CPR [see Eq.\ \eqref{eq: sharp drops}] look almost vertical. At the same time, the not-so-steep linear segments are the parts of the nanobridge CPR [see Eq.\ \eqref{eq: smooth segments}]. (b)~Dependence of the critical current asymmetry $\Delta I_{c}$ on the magnetic field. (c)~Dependence of the phases $\varphi_{\mathrm{max}}$ corresponding to the critical current, on the magnetic field for positive (red) and negative (blue) current directions. The steep (almost vertical) segments in panel (a) correspond to $\varphi_{\mathrm{max}} = \pm \varphi_{c}$ [flat parts of $\varphi_{\mathrm{max}}(B)$], while the not-so-steep linear segments correspond to the linear parts of $\varphi_{\mathrm{max}} (B)$. The transition between the two regimes of $\varphi_{\mathrm{max}} (B)$ is abrupt.  The parameters used for plotting the curves are taken from Table \ref{table: samples} for sample N7.}
\label{fig:critical currents}
\end{center}
\end{figure*}

{When critical currents are reached at the $\pm \varphi_{c}$, they take the values determined by Eq.\ \eqref{eq: sharp drops}. This behavior corresponds to the steep (almost vertical) segments in the $I_{c\pm}(B)$ dependence in Fig.~\ref{fig:critical currents}(a) [which is the reproduced theoretical curves for the highest branches of the critical currents from Fig.~\ref{fig: Ic(H)}(b)]. However, this is nothing more than part of the sinusoidal CPR of the SNS junction.}

{At the same time, when critical currents are reached at $\varphi^{*}_{\pm}$, they are given by Eq.\ \eqref{eq: smooth segments}. This behavior corresponds to the not-so-steep linear segments in Fig.~\ref{fig:critical currents}(a), and this is nothing more than a part of the linear CPR of the nanobridge.} 

Note that $\varphi^{*}_{\pm}$ exists only when the critical current of the SNS junction is large enough so that  $\kappa = I_{\mathrm{sns}}(\phi) \varphi_{c}/I_{\mathrm{nb}} > 1$. At the same time, at large magnetic fields, this condition is violated because of Fraunhofer suppression of the SNS critical current. Therefore, in this regime, critical currents are reached at $\pm \varphi_{c}$. However, we do not conduct our measurements in such strong magnetic fields and do not observe such a phenomenon. 

Additionally, we examine the asymmetry of the critical current, defined as $\Delta I_{c}(\phi) = I_{c+}(\phi) - I_{c-}(\phi)$. In the regime where both critical currents are reached at $\pm \varphi_{c}$, we find
\begin{equation}\label{eq:trivial asymmetry}
    \Delta I_{c}= 2 I_{\mathrm{sns}} (\phi) \cos{\varphi_c}\sin{\phi}.
 \end{equation}
In the other regime, when both critical currents are reached at $\varphi_{\pm}^{*}$, we obtain { 
\begin{equation}\label{eq:nontrivial asymmetry}
     \Delta I_{c} =  \frac{2 I_{\mathrm{nb}}}{\varphi_{c}}\left[ \pi \left(N_{v+} - N_{v-}\right) - \phi \right].
 \end{equation}}
Finally, it is possible that one critical current is reached at $\pm \varphi_c$, while the second one is reached at $\varphi^{*}_{\pm}$. The results are illustrated in Figs.~\ref{fig:critical currents}(b) and~(c). 
 
\section{Temperature dependence of junction parameters}\label{appendix:temperature dependence of junction parameters}

We approximate the experimentally measured critical currents $I_{\mathrm{nb}}$, $I_{\mathrm{sns}}$, and critical phase $\varphi_c$ dependencies on the temperature $T$ in the following way.  For the nanobridge, we assume the Likharev-Yakobson model \cite{LikharevYakobson1975, LikharevRevModPhys.51.101, GolubovRevModPhys.76.411} formulated for long and narrow superconducting bridges. In the framework of this model, the critical current and critical phase have the following temperature dependencies,
\begin{gather}
I_{\mathrm{nb}} (T) = I^{(f)}_{\mathrm{nb}} (1 - T/T_{\mathrm{tr}})^{3/2}, \\ \varphi_{c}(T) = \varphi^{(f)}_{c} \sqrt{1 - T/T_{\mathrm{tr}}},
\end{gather}
where we take $I^{(f)}_{\mathrm{nb}} = 88$\,$\mu$A, $\varphi^{(f)}_{c} = 62$, and $T_{\mathrm{tr}} = 2.4$\,K as fitting parameters. Here, $T_{\mathrm{tr}}$ is the transition (critical) temperature of the nanobridge, which is significantly smaller than the critical temperature of the thick niobium, as mentioned in Sec.\ \ref{sec: critical current measurements}. The results of our approximation are presented in Figs.~\ref{fig: Ic(H)}(e) and~(f), respectively.

At the same time, we consider the junction formed by the flake of Bi${}_2$Te${}_2$Se as a long dirty SNS junction, with the critical current given by the expression from Refs. \cite{Dubos2001PhysRevB,ZaikinZharkov1981},
\begin{equation}
 I_{\mathrm{sns}}(T) = \frac{64 \pi T}{e R_{\mathrm{sns}}} \sum_{n=0}^{\infty} \frac{L}{L_{\omega_n}} \frac{\Delta^2 \exp{(- L/L_{\omega_n})}}{\left[\omega_n + \Omega_n + \sqrt{2( \Omega_{n}^2 + \omega_n \Omega_n) }\right]^2}, 
\end{equation}
where $\omega_{n} = (2n +1) \pi T$, $\Omega_{n} = \sqrt{\omega_{n}^2 + \Delta^2}$, and $L_{\omega_{n}} = \sqrt{\hbar D/2\omega_{n}}$. 

Here, $R_{\mathrm{sns}}$ is the effective resistance of the SNS junction, $L$ is the distance between the superconducting electrodes, $D$ is the diffusion constant of the N metal, and $\Delta$ is the proximity-induced superconducting gap of the regions beneath the flake. Experimentally, the critical temperature of the SNS junction is found to be approximately equal to the transition temperature of the nanobridge. Therefore, for simplicity, we assume that the SNS junction has the same critical temperature $T_{\mathrm{tr}}$ (which enters $\Delta$) as the nanobridge. We take $R_{\mathrm{sns}}$ and the relation between the junction length $L$ and the thermal length $L_{T} = \sqrt{\hbar D / 2 \pi T}$ at $T = 2$~K as fitting parameters. To obtain the results presented in Fig.~\ref{fig: Ic(H)}(d), we choose the following values: $R_{\mathrm{sns}} = 15.5$~$\Omega$, and $L/L_{T} = 5.73$ at $T= 2$~K. In the relevant temperature range, $T \sim (1-3)$~K, the assumption that the SNS junction is long is satisfied since $L/L_{T} \sim 5$.

\section{Shapiro steps shift}\label{appendix:Shapiro steps shift}

The reasons for the Shapiro steps to be located at half-integer values in the case of the discussed sample (N7) remain unclear. During this research, three samples were thoroughly measured, and the additional data can be found in the Appendix~\ref{sec: More Shapiro steps data}. First of all, the phenomena discussed in this paper are not unique to a single sample, since the same shift and diode effect in Shapiro steps were observed in another SQUID (N8) with similar parameters (see Table~\ref{table: samples}), as can be seen in Figs.~\ref{fig: Appendix_moredata}(a) and (b). At the same time, both effects are absent in data from sample N5, see Figs.~\ref{fig: Appendix_moredata}(c) and (d). We attribute this to the fact that the critical current through the SNS junction in this sample is an order of magnitude smaller than in two other samples and than the critical current of the nanobridge (see Table~\ref{table: samples}). This means that in the R state, because of overheating, the SNS critical current becomes negligible, and the SQUID works as a nanobridge shunted by normal resistance. Summarizing all the above, we observe shifted steps in samples N7 and N8, and unshifted ones in sample N5. This eliminates the possibility of errors because of the measuring system or b data processing.

We measured half-integer Shapiro steps at different frequencies in the range from $f_{\mathrm{rf}}= 2.8$\,GHz to $f_{\mathrm{rf}}=3.8$\,GHz; the results are demonstrated in Fig.~\ref{fig: Appendix_moredata}(e). At the same time, Fig.~\ref{fig: Appendix_moredata}(f) shows that at high rf power, when the supercurrent is fully suppressed, we are able to observe Shapiro steps from $|V|/V_0 = \frac{1}{2}$ to $|V|/V_0 =17\frac{1}{2}$. Thus, the shifted position of the Shapiro steps cannot be related to a series-connected resistance since it would change the voltage period of the steps, which would be noticeable in such a wide range of voltages.

At the same time, we note that the shift of the Shapiro features cannot be described theoretically by our RSJ model presented by Eqs.~\eqref{eq:J1beg} and \eqref{eq:J2beg}, which predicts positions of the steps to be exactly at $\overline{v} = (n/k) \omega$ (where  $n$ and $k$ are integers) with the most pronounced steps corresponding to $k = 1$. Therefore, we must consider additional factors beyond the presented RSJ model, which may be related to the complex dynamics within the junctions themselves or in the SQUID.

One of these factors is geometrical inductance $L_{\mathrm{loop}}$ of the SQUID, which we neglect in our model. For example, Ref.\ \cite{Half_steps_in_SQUID_Early1995} shows that even if both JJs in SQUID have a sinusoidal CPR, with a large enough $L_\mathrm{loop}$, fractional steps may occur because of their sequential phase changing. Such behavior was demonstrated experimentally in Ref.\ \cite{Ciaccia2024}, where the half step became even larger than the first one at specific values of magnetic flux. However, this effect disappears when the SQUID contains an integer number of flux quanta. In contrast, in our case, the Shapiro steps are at half-integer positions at arbitrary flux, so this mechanism is not relevant in our case. In addition, our estimations presented in Appendix~\ref{appendix:geometric inductance} show that the influence of inductance is negligible in our case.

Another possible factor is connected with the properties of the R state in the nanobridge. Our nanobridge is long (in units of superconducting coherence length), and it may contain several intrinsic ``weak'' places along its length. At $I>I_{\mathrm{nb}}$ the resistive state starts from vortex penetration via the weakest place and vortex passage through the bridge. As the current increases, the vortices may start entering other weak places and interact with vortices nucleated in other places. The overall behavior can be rather complicated; see, for example, Ref.\ \cite{Grebenchuk_2022}, where the response of several weak places connected in series to rf irradiation has been studied. 

The relevance of these factors to the explanation of the Shapiro steps shift is not clear at present. Moreover, there may be other factors that could cause this effect and which are not mentioned above. For example, a similar shift of the Shapiro steps was observed in Ref.\ \cite{Kazmin2024} in a completely different system for unknown reasons. 

\section{{Comparison of the theoretical results for the Shapiro features at different temperatures}}\label{appendix: comparison of the results for high and low temperatures}

{As discussed in Sec.~\ref{sec:Shapiro measurments}, the theoretical results in Fig.~\ref{fig: Shapiro_5} show only qualitative agreement with the experimental data. While the experimental data exhibit three distinct features in the shown voltage range, our theoretical model reproduces only two of them. The third feature (corresponding to $n = \pm 9$) is predicted to be comparable in magnitude to the thermal noise level, rendering it effectively unobservable. In addition, the depths of the Shapiro features decrease much faster with the voltage in theory compared to the experiment.}

{The most straightforward explanation for this discrepancy would be an overestimation of the effective temperature and its associated thermal fluctuations. However, both of our estimates of effective temperature, based on $I_{r}$ measurements and experiments conducted under external irradiation, lead to the result that $T \sim 2$~K, which aligns with the temperature value used in our calculations presented in Fig.~\ref{fig: Shapiro_5}.} 

\begin{figure*}[ht!]
\begin{center}
\includegraphics[width=17.5cm]{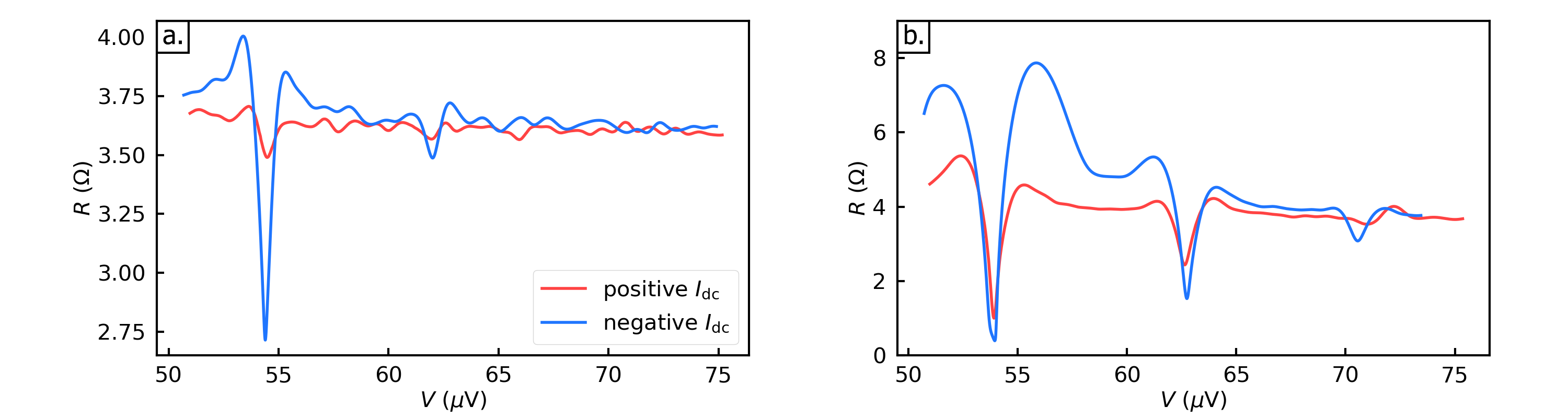}
\caption{{Theoretical results within the RSJ model at $\Phi/\Phi_0 = 1/4$, $I_{\mathrm{ac}} = 12\,\mu$A, and different temperatures taking into account the temperature dependencies of junction parameters: (a) $T= 1.8$\,K, $\varphi_{c} = 30$, $I_{\mathrm{sns}} = 0.3\,\mu$A, and $I_{\mathrm{nb}} = 12\,\mu$A, (b) $T = 1$\,K, $\varphi_{c} = 47$, $I_{\mathrm{sns}} = 2.8\,\mu$A, $I_{\mathrm{nb}} = 37\,\mu$A. As temperature decreases, both the depths and widths of the Shapiro features increase, which is the result of a decrease in the strength of thermal fluctuations.}}
\label{fig:Additional plot for Shapiro step}
\end{center}
\end{figure*}

{At the same time, by allowing the temperature to take arbitrary values within the model, we can generate plots similar to those in Fig.~\ref{fig: Shapiro_5}, albeit with varying sizes of the Shapiro features. For example, the comparison of the results for the Shapiro features at $T = 1.8$\,K and $T = 1$\,K are shown in Fig.~\ref{fig:Additional plot for Shapiro step}. It is evident that both the depth and width of the Shapiro features increase as the temperature decreases. Furthermore, because of the reduction of thermal noise effects, it becomes feasible to observe the  $n = \pm 9$  Shapiro features in Fig.~\ref{fig:Additional plot for Shapiro step}(b). However, despite the qualitative similarity of the theoretical curves to those observed experimentally, significant discrepancies remain. As illustrated in Fig.~\ref{fig:Additional plot for Shapiro step}, the depths of the theoretically calculated Shapiro features are considerably larger in amplitude compared to the experimental results. Additionally, there remains an issue regarding the rapid decrease in the depth of the Shapiro features with increasing voltage.}

{Considering all the mentioned-above factors, we have decided to use the results shown in the theoretical curve in Fig.~\ref{fig: Shapiro_5}, which are calculated at  $T = 1.8$~K, for comparison with the experimental data in the main text.}

\section{Experimental methods}
\label{sec: Experimental methods}
The measurements were carried out in a dilution refrigerator at a temperature $T = 50$~mK. IV curves of the SQUID were measured using the four-terminal technique. For the measurements of the critical current oscillations, we applied an \emph{ac} current with the frequency 17\,Hz using a Low-distortion Function Generator SRS DS360. The voltage response was first amplified and filtered with a preamplifier SRS SR560, and then digitized by the NI-9234 ADC. The critical current was identified when the voltage exceeded the threshold value $V_{\text{thr}} = 20\,\mu$V. 
During the measurements of the differential resistance, we used a different setup with a \emph{dc} current source Keithley 6221 and a nanovoltmeter Keithley 2182A. The rf signal was generated using VNA Cobalt C1220 and delivered to the sample through a coaxial cable with a $-20$\,dB cryogenic attenuator. The open end of the cable was fixed to the holder a few millimeters above the surface of the sample, acting as an $rf$ antenna.

To reduce the influence of noise effects on the experimentally obtained dependencies presented in Fig.~\ref{fig: Shapiro_analysis}, we average the Shapiro step depths in the following way. As $\Delta R_{\pm}$ we take the average value of the resistance in the  voltage range $[V_{n}/V_0 - 1/2,V_{n}/V_0+1/2]$, where $V_{n}$ is the voltage corresponding to the $n$th Shapiro feature center. This procedure leads to smoothing of the curves presented in Fig.~\ref{fig: Shapiro_analysis}, while decreasing the depth of the steps. It results in a difference in the step depths, presented in the experimental panels in Figs.~\ref{fig: Shapiro_5} and~\ref{fig: Shapiro_analysis}.

\section{Method of slowly varying phase and effective Josephson equations}\label{appendix: effective Josephson equations}
As mentioned in Sec.\ \ref{sec:Analysis of the Shapiro features}, to find the shapes of the $n$th  Shapiro features (with both positive and negative numbers) in the vicinity of their centers, we apply the first iteration of the perturbation method of slowly varying phase. Specifically, we represent the phase as a sum of ``fast'' [varies at short time scale $\tau \lesssim (n \omega)^{-1}$] and ``slow'' [varies at long time scale $\tau  \gg (n \omega)^{-1}$] parts, respectively, $\varphi(\tau) = \varphi_{f} (\tau) + \varphi_{s} (\tau)$. We assume that supercurrent and thermal fluctuations are small compared to \emph{dc} current [accurate conditions of applicability are discussed below Eq.\ \eqref{eq:SC}]. Particularly, the CVC is close to Ohm's law $\hat{v} = n \omega$. At the same time, the fast dynamics of the phase is determined by \emph{dc} and \emph{ac} currents, while the supercurrent and thermal fluctuations are responsible only for the slow corrections to the phase dynamics. Therefore, in the zeroth order of the perturbation theory
\begin{gather}\label{eq:varphi0}     \varphi_{f}(\tau) =  n \omega \tau + (j_{\mathrm{ac}}/\omega) \sin{\omega \tau}, \quad \varphi_{s}(\tau) = \theta_{n}(\tau). 
\end{gather}
It is important to remember that even though $\theta_{n}(\tau)$ is a slow function, it contains a linearly growing with time part, which corresponds to the slow part of $\hat{v}(\tau)$, see Eq.\ \eqref{eq:slow variables}.

In the first order, we take the supercurrent and thermal fluctuations into account. For this, we consider $\xi(\tau)$  in the right-hand side of Eq.\ \eqref{eq:J1beg} and also substitute $\varphi_f (\tau)$ to $I_{s}(\varphi)$ and expand the resulting expression into the Fourier series:
\begin{multline}\label{eq:SC before averaging}
I_{s\pm}(\tau) = \sum \limits_{k=1}^{\infty} \sum \limits^{\infty}_{m=-\infty} (-1)^{m} A_{k\pm} J_{m}\left(\frac{k j_{\mathrm{ac}}}{\omega}\right) \\
\times \sin{\bigl(k (\theta_n + \delta_{k\pm}) + (k n - m) \omega \tau \bigl)}.
\end{multline}
After that, we average the Josephson equation in the first order with supercurrent given by Eq.\ \eqref{eq:SC before averaging} over time $\Delta \tau$ such that $(n \omega)^{-1} \ll \Delta \tau \ll {\cal{T}}^{-1}, (\hat{v} - n \omega)^{-1} $ to exclude fast oscillating terms and obtain the effective equation for the slow functions. Averaging over $\Delta \tau$ implies: (i)~keeping $\theta_{n}$ intact in the left-hand side of Eq.\ \eqref{eq:J1beg}, (ii)~keeping only the slow part of $\xi (\tau)$ (which in this order of perturbation theory can still be considered as a white noise), and (iii)~keeping in the supercurrent Eq.\ \eqref{eq:SC before averaging} only terms with $m = k n$ while neglecting all other terms since they oscillate with frequencies $\gtrsim \omega$. As a result of this procedure, we obtain the effective Josephson equation in the form of Eq.\ \eqref{eq:effective} with the averaged supercurrents given by  
\begin{equation}\label{eq:SC}
\hat{I}_{s\pm}(\tau) = \sum \limits_{k=1}^{\infty}  (-1)^{nk} A_{k\pm} J_{nk}\left(\frac{k j_{\mathrm{ac}}}{\omega}\right)\sin{\Bigl(k[\theta_{n}(\tau)+ \delta_{k\pm}]\Bigl)}.
\end{equation}
One of the validity regimes of Eq.\ \eqref{eq:effective}, which is relevant for our case, is $\Delta j \ll 1$ (small deviation from the center of noise-free Shapiro steps),  $A_{k} \ll 1$ (small supercurrent compared to dc), ${\cal{T}}/ n \omega \ll 1$ (small thermal fluctuations compared to \emph{dc} currents contributions).

At the same time, if we neglect harmonics with $k>1$ and make the phase shift $\theta_{n} \mapsto \theta_{n} - \delta_{1\pm} - \pi n$ (which does not change the CVC), we obtain the supercurrent in the single-harmonic approximation given by Eq.\ \eqref{eq:Is pm}. The single-harmonic approximation is rigorously justified in the limiting case $j_{\mathrm{ac}}/\omega \ll 1$, when higher harmonics in Eq.\ \eqref{eq:SC} are parametrically smaller than the first one.

\section{Calculation of \texorpdfstring{$R_{\pm}$}{R+-} in the single-harmonic approximation} \label{sec: CVC in the single harmonic approximation}

As mentioned in Sec.~\ref{sec:Analysis of the Shapiro features}, in the single-harmonic approximation [Eq.\ \eqref{eq:Is pm}], effective Josephson Eq.\  \eqref{eq:effective} has the form of the ordinary Josephson equation for the phase $\theta_{n}$ with sinusoidal CPR:
\begin{gather}\label{eq:J1term}
     \dot{\theta}_{n} + j_{1\pm} \sin \theta_{n} =  \Delta j + \xi(\tau),  \\ j_{1\pm} = \left|A_{1\pm} J_{n}(j_{\mathrm{ac}}/\omega)\right| \notag.
\end{gather}
For further analysis, it is convenient to work with the stationary Fokker-Planck equation on the distribution functions $P_{\mathrm{st}\pm}(\theta_n)$ which corresponds to the Langevin equation \eqref{eq:J1term}:
\begin{gather}\label{eq:FP}
\frac{\partial P_{\mathrm{st}\pm}}{\partial \theta_{n}} + \frac{1}{\cal{T}_{\pm}}\left(\sin{\theta_n} - \Delta j/j_{1\pm}\right)P_{\mathrm{st}\pm} = Q_{\pm},
\end{gather}
where $Q_{\pm}$ is the quantity that should be found from the formalization condition on $P_{\mathrm{st}\pm}(\theta_n)$.

To obtain the CVC, one needs to solve the Fokker-Planck equation in the form of Eq.\ \eqref{eq:FP} and calculate the statistically-averaged voltage. For definiteness, we consider $\Delta j >0$ and take into account the symmetry of the Shapiro features at the end of the calculation. In this way, we obtain the voltage contribution from the slow dynamics, which is added to $n \omega$. As a result
\begin{multline}\label{eq:votage} 
 |\langle \overline{v} \rangle - n \omega |= \int \overline{\dot{\varphi}} P_{\mathrm{st}\pm}(\varphi) d\varphi  \\ = \int (\Delta j - j_{1\pm}\sin \varphi) P_{\mathrm{st}\pm}(\varphi) d\varphi  =   -2 \pi j_{1\pm} { \cal{T_{\pm}} } Q_{\pm}.
\end{multline}
We now only need to calculate $Q_{\pm}$. It can be done with the use of the variation of parameters method taking into account the normalization condition and periodicity of $P_{\mathrm{st}\pm}(\varphi)$,
\begin{multline}\label{eq:Qpm}
  \frac{1}{Q_{\pm}}= - \int 
\limits^{2 \pi}_{0} d\varphi \exp\left( \frac{\varphi \Delta j /j_{1\pm} + \cos \varphi}{\cal{T_{\pm}}}\right) \\
\times \int \limits^{\infty}_{\varphi}dx \exp\left(-\frac{x \Delta j /j_{1\pm}  + \cos x}{\cal{T_{\pm}}}\right).   
\end{multline}

To calculate the integrals in Eq.\ \eqref{eq:Qpm}, we perform the following Fourier expansion: 
\begin{equation}\label{eq: Qpm Fouirer}
    \exp\left( \frac{\cos \varphi}{\cal{T_{\pm}}}\right) = I_{0}\left(\frac{1}{\cal{T_\pm}}\right) + 2 \sum \limits_{k=1}^{\infty} I_{k}\left(\frac{1}{\cal{T_\pm}}\right) \cos k \varphi.
\end{equation}
Using Eq.\ \eqref{eq: Qpm Fouirer}, we can represent $Q_{\pm}$ in the following form:
\begin{gather}
   \frac{1}{Q_\pm} = -\int \limits_{0}^{2 \pi} d \varphi \left[ I_{0}\left(\frac{1}{\cal{T_\pm}}\right) + 2 \sum \limits_{k=1}^{\infty} I_{k}\left(\frac{1}{\cal{T_\pm}}\right) \cos k  \varphi\right]  \notag \\ 
   \times \biggl[ I_{0}\left(\frac{1}{\cal{T_\pm}}\right) \frac{\Delta j}{j_{1 \pm} {\cal{T_\pm}}} + \sum \limits_{m=1}^{\infty} I_{m} \left(\frac{1}{\cal{T_\pm}}\right) \frac{2  (-1)^m}{m^2 + (\Delta j/ j_{1 \pm} {\cal{T_\pm}})^2} \notag \\ \times \left( \frac{\Delta j}{j_{1 \pm} {\cal{T_\pm}}} \cos m \varphi - m \sin m \varphi \right)\biggl]. \label{eq:Qpm answer}
\end{gather}
Performing the integration over $\varphi$ and substituting Eq.\ \eqref{eq:Qpm answer}  to Eq.\ \eqref{eq:votage}, we obtain the voltage:
\begin{equation}\label{eq:CVC general}
 \langle \overline{v}\rangle= n \omega  + \Delta j \biggl[ \sum^{\infty}_{k=0}
  I^2_{k}\left(\frac{1}{{\cal{T_{\pm}}}}\right) \frac{(-1)^k (2 - \delta_{k, 0})\left(\Delta j/ j_{1\pm}\right)^2}{(\Delta j / j_{1\pm})^2 + {\cal{T_{\pm}}}^2 k^2}\biggl]^{-1},
\end{equation}
which is valid for both signs of $\langle\overline{v}\rangle$ and $\Delta j$.
Taking the derivative of Eq.\ \eqref{eq:CVC general}, we obtain the expression for the differential resistance Eq.\ \eqref{eq:R_diff} from the main text.

\section{Dependence of the Shapiro steps on power of microwave irradiation}\label{appendix: dependence on Pac}

\begin{figure*}[ht!]
\begin{center}
\includegraphics[width=17.5cm]{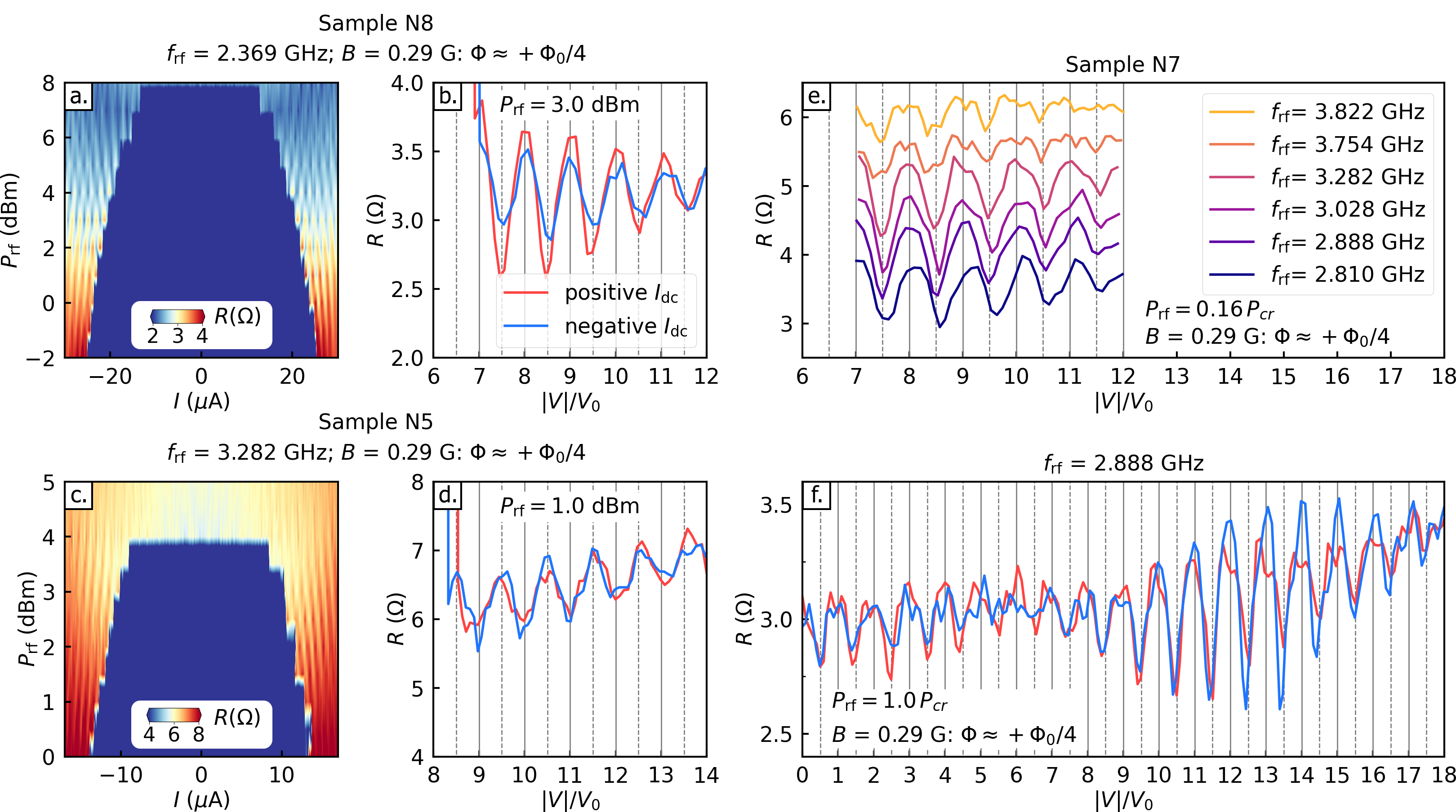}
\caption{Additional data of the Shapiro steps measurements for different samples. {Panel (a) shows the dependence of the differential resistance on the backward branches near $I_{r\pm}$ vs current through the sample and rf power in the external magnetic field $B = 0.29$\,G for sample N8. Panel (b) shows the Shapiro features in the positive and negative \emph{dc} current directions under the same conditions.} The dips of the $R(V)$ dependence are located at $|V|/V_0 = n - 1/2$ with integer $n$. The plots clearly show the asymmetry of the Shapiro features for different current directions. 
Vertical left segments of the curves correspond to the abrupt retrapping to the S state. Panels (c) and (d) present similar to (a) and (b) dependencies for sample N5 at the same magnetic field. However, in this sample, the Shapiro features are located at integer values of $|V|/V_0$ and do not demonstrate noticeable asymmetry. 
(e)~Curves for sample N7 at different frequencies of the external microwave signal. An additional shift in 0.5~$\Omega$ was added sequentially to the data at different frequencies. The Shapiro features are located at half-integer positions for all frequencies. $P_{\mathrm{cr}}$ is defined as the minimal rf power at which the retrapping current of the sample $I_{r}$ is fully suppressed. (f) Shapiro features in sample N7 measured at high microwave power, when the retrapping current is fully suppressed. At this condition, we are able to observe 18 dips in the differential resistance, all located at half-integer values of $|V|/V_0$. With the increase of $|V|/V_0$, the diode effect changes its sign. We attribute this to the Joule heating effect caused by \emph{dc} current and temperature dependence of the critical phase $\varphi_{c}$. At the same time, an unexpected increase of the Shapiro features depths is observed in the range of $|V|/V_0$ between $8$ and $13$.}
\label{fig: Appendix_moredata}
\end{center}
\end{figure*}

As mentioned in Sec.~\ref{sec:Analysis of the Shapiro features}, the dependence of the Shapiro feature depths on the power of \emph{ac} irradiation demonstrates nonmonotonic behavior caused by the Joule heating effect. Our goal is to explain such behavior and obtain theoretically the dependence of differential resistance on rf irradiation power, which would reproduce the experimental result of Fig.~\ref{fig: Shapiro_analysis}(c) measured at $\Phi/\Phi_0 = 0.75$ for $n = \pm 7$. To do that, we need to relate the power of rf irradiation $P_{\mathrm{rf}}$ and the temperature of the electron system $T$ and then substitute the $T(P_{\mathrm{rf}})$ dependence to Eq.\ \eqref{eq:R_diff}. Moreover, we need to relate the power of rf irradiation and the \emph{ac} current, $I_{\mathrm{ac}}(P_{\mathrm{rf}})$, which also enters Eq.\ \eqref{eq:R_diff}.

We use the energy balance equation assuming that the electron system is cooled by the phonon subsystem \cite{Cecco2016},
\begin{equation}\label{eq:energy balance}
P_{\mathrm{rf}}/Y + P_{\mathrm{dc}} = C_{\text{e-ph}} \left( T^5 - T^{5}_{0} \right),
\end{equation}
where $Y$ is the factor determining what part of external microwave irradiation power is transferred to the sample, $C_{\text{e-ph}}$ is the coefficient of the electron-phonon coupling, $T_{0} = 50$~mK is the temperature of the phonon subsystem, and $P_{\mathrm{dc}} = I_{\mathrm{dc}}^2 R_0$ is the Joule heating power owing \emph{dc} current $|I_{\mathrm{dc}}| = 20$\,$\mu$A, which corresponds to the Shapiro steps centers. We assume that the \emph{ac} current is produced by microwave irradiation and thus
\begin{equation}\label{eq:ac current}
    P_{\mathrm{rf}}/Y = I_{\mathrm{ac}}^2 R_{0}/2.
\end{equation}
For the fitting procedure, we rewrite Eq.\ \eqref{eq:energy balance} in the form
\begin{equation}\label{eq: energy balance fit}
    T = A_{\mathrm{fit}} (P_{\mathrm{rf}} + Y P_{\mathrm{dc}})^{1/5},
\end{equation}
where $A_{\mathrm{fit}} = (C_{\text{e-ph}} Y)^{-1/5}$.
Here, we neglect $C T_0^5$ term because of its small value.   

The $R_{\pm}(P_{\mathrm{rf}})$ dependence can be numerically obtained from Eqs. \eqref{eq:ac current} and \eqref{eq: energy balance fit}. 
In Eq. \eqref{eq:ac current} we use the value of  \emph{ac}  current $I_{\mathrm{ac}}$ at some value of rf power $P_{\mathrm{rf}}$ as the fitting parameter (instead of $Y$). For this purpose, we choose $P_{\mathrm{rf}} = 1$~mW because it corresponds to the location of the most pronounced maximum in Fig.~\ref{fig: Shapiro_analysis}(c).
At the same time, in Eq.\eqref{eq: energy balance fit} we use $A_{\mathrm{fit}}$ as the second fitting parameter.  

The fitting procedure generates plots similar to Fig.~\ref{fig: Shapiro_analysis}(f) with a different number of extrema, their amplitudes, and locations. In our fitting procedure we require the location of the highest  maximum to be at $P_{\mathrm{rf}} = 1$\,mW, as it is in Fig.~\ref{fig: Shapiro_analysis}(c). We also adjust the number of extrema and their amplitudes in theory with those in Fig.~\ref{fig: Shapiro_analysis}(c). As a result, the fitting parameters are $A_{\mathrm{fit}} = 1.72$\,K/mW$^{1/5}$, and $I_{\mathrm{ac}} (1\,\mathrm{mW}) = 15.7$\,$\mu$A. At the same time, our model predicts that temperature changes from $2.2$~K at $P_{\mathrm{rf}} = 0$\,mW to $2.4$\,K at $P_{\mathrm{rf}} = 2$\,mW.
The result is presented in Fig.~\ref{fig: Shapiro_analysis}(f).

\section{Additional Shapiro steps data}
\label{sec: More Shapiro steps data}

\begin{table}[ht!]
\caption{Parameters of the three measured SQUID samples at $T = 50$~mK: the critical current of the SNS junction $I_{\mathrm{sns}}$, the nanobridge critical current $I_\mathrm{nb}$, the critical phase $\varphi_c$, and the normal resistance $R_0$ of the whole SQUID.}
\begin{center}
\begin{tabular}{c c c c c}
Sample  & $I_\mathrm{sns}$ ($\mu$A)   & $I_\mathrm{nb}$ ($\mu$A)    & $\varphi_c$   & $R_0$ ($\Omega$)    \\ \hline
N5      & 1.6                   & 48.9                  & 63.8          & 7.5               \\ 
N7      & 23.0                  & 53.7                  & 58.5          & 3.6               \\ 
N8      & 11.1                  & 62.9                  & 61.5          & 2.8               \\  
\end{tabular}
\end{center}
\label{table: samples}
\end{table}

During the experiments, we measured three samples (N5, N7, and N8) with the same SQUID and nanobridge designs but different BTS flakes serving as the SNS junction. Sample N7 is discussed in the main text. All SQUIDs were studied under a magnetic field at 50~mK. Using Eq.\ \eqref{eq:supercurrent} to fit experimentally measured curves $I_c(H)$, we determined the main parameters of the samples, listed in Table~\ref{table: samples}. 

For all samples, the CVCs were measured at various microwave frequencies and powers under different magnetic fields. Figure~\ref{fig: Appendix_moredata} shows some examples of the obtained results discussed in  Appendix~\ref{appendix:Shapiro steps shift}.

\bibliography{references}
\end{document}